\documentclass[conference]{acmart}

\usepackage[T1]{fontenc}
\usepackage{graphicx}
\usepackage{amsmath}
\usepackage{amsfonts}
\usepackage{graphicx}
\usepackage{tikz}
\usepackage{siunitx}
\usepackage{multirow}
\usepackage{booktabs}
\usepackage{lipsum}
\usepackage{pifont}
\usepackage{threeparttable}
\usepackage{subcaption} 
\usepackage[ruled,vlined,boxed]{algorithm2e}
\title{An MLIR-Based Compilation Framework for Control Flow Management on Coarse Grained Reconfigurable Arrays}

\usepackage{xcolor}
\definecolor{darkgreen}{rgb}{0.0, 0.5, 0.0}  
\usepackage{flushend}

\usepackage{listings}

\begin{abstract}
Coarse-Grained Reconfigurable Arrays (CGRAs) present both high flexibility and efficiency, making them well-suited for the acceleration of intensive workloads. Nevertheless, a key barrier towards their widespread adoption is posed by CGRA compilation, which must cope with a multi-dimensional space spanning both the spatial and the temporal domains. Indeed, state-of-the-art compilers are limited in scope, as they mostly handle data flow while providing little or no support for control flow. Hence, they mostly focus on mapping single loops and/or delegate the management of control-flow divergences to ad-hoc hardware units. 

Conversely, in this paper, we show that control flow can be effectively managed and optimized at the compilation level, enabling a broad set of applications to be targeted while remaining hardware-agnostic and achieving high performance. We embody our methodology in a modular compilation framework consisting of transformation and optimization passes, enabling support for applications with arbitrary control flows running on {homogeneous} CGRA meshes. We also introduce a novel mapping methodology that serves as a compilation back-end, addressing limitations in available CGRA hardware resources and ensuring a feasible solution during compilation.
Our framework achieves up to 2.1$\times$ {runtime} speed-up over state-of-the-art approaches, purely through compilation optimizations.
\end{abstract}

\author{Yuxuan Wang}
\email{yuxuan.wang@epfl.ch}
\affiliation{%
  \institution{EPFL}
  \city{Lausanne}
  \country{Switzerland}
}

\author{Cristian Tirelli}
\email{cristian.tirelli@usi.ch}
\affiliation{%
  \institution{USI}
  \city{Lugano}
  \country{Switzerland}
}

\author{Giovanni Ansaloni}
\email{giovanni.ansaloni@epfl.ch}
\affiliation{%
  \institution{EPFL}
  \city{Lausanne}
  \country{Switzerland}
}

\author{Laura Pozzi}
\email{laura.pozzi@usi.ch}
\affiliation{%
  \institution{USI}
  \city{Lugano}
  \country{Switzerland}
}

\author{David Atienza}
\email{david.atienza@epfl.ch}
\affiliation{%
  \institution{EPFL}
  \city{Lausanne}
  \country{Switzerland}
}

\begin{document}
\maketitle

\section{Introduction}

Coarse-Grained Reconfigurable Arrays (CGRAs) are a spatial architecture composed of Processing Elements (PEs) connected in a mesh topology~\cite{839320, adres}. PEs comprise ALUs and local registers, supporting arithmetic operations. Being programmable at the operation level, CGRAs present far lower overhead and much faster reconfiguration time with respect to FPGAs.  
Indeed, as opposed to mainstream FPGAs, CGRA{s} typically host multiple configurations, one of which is selected at run-time for each clock cycle.
This reconfiguration capability allows CGRAs to  effectively support the acceleration of intensive parts of applications (computational kernels) defined in high-level languages (e.g., C/C++) through spatial mapping and temporal configuration. 
CGRAs provide a high degree of flexibility and performance, attributed to their reconfigurability and data-parallel execution model, respectively~\cite{cgra_me}.

\begin{figure}[t!]
    \centering
    \includegraphics[width=0.9\linewidth]{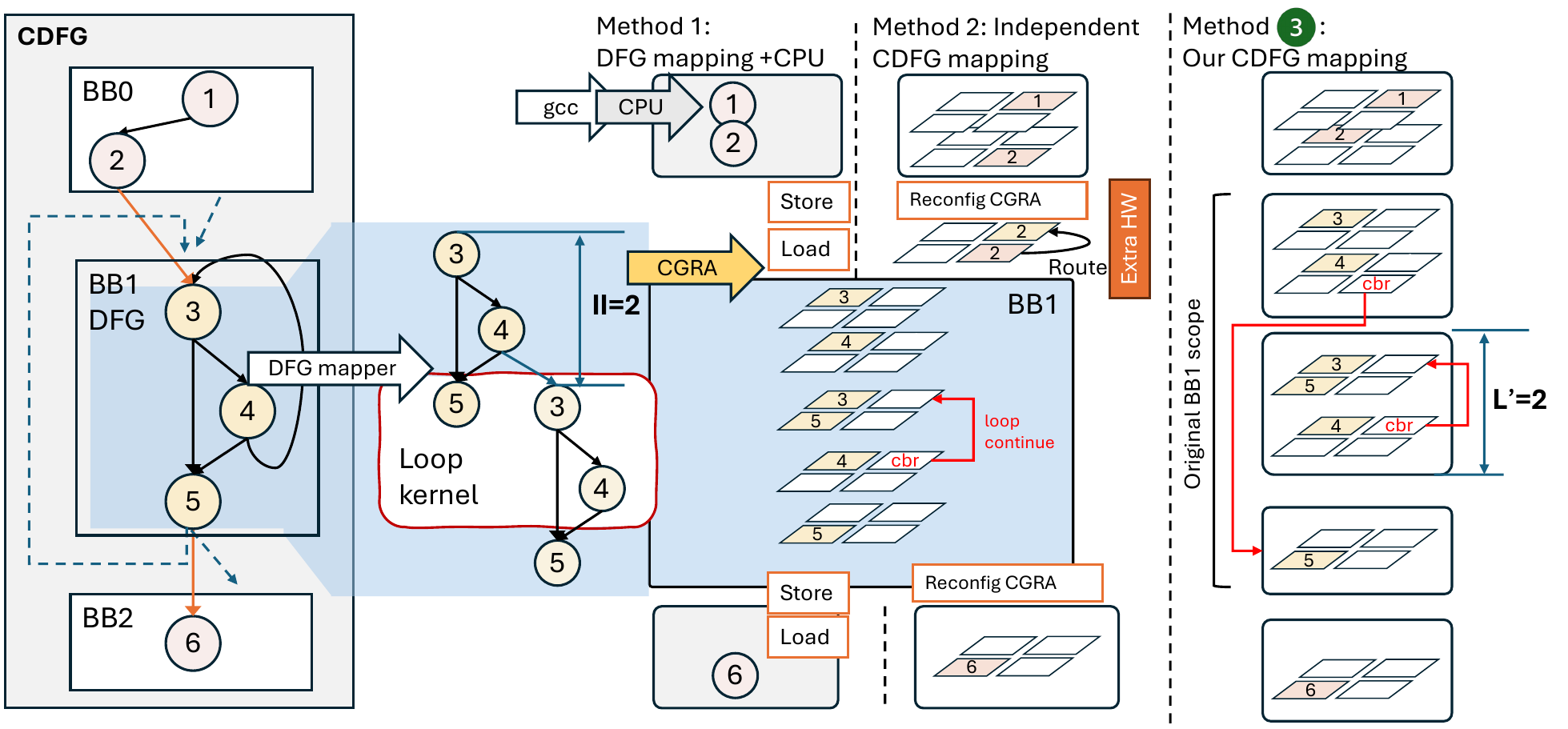}
    \caption{Compiling applications onto the CGRA hardware. 
    Method 1: DFG mappers fold loops to exploit parallelism, while offloading operations outside their scope to the host processor. This approach requires frequent data exchange between CGRA and host.
    Method 2: Independent CDFG mapping treats each basic block as an isolated DFG during compilation. Data and instruction reconfiguration are managed through dedicated hardware.
    Method 3: Our mapper targets CDFG compilation with branches for CFG management, resulting in no reconfiguration overhead for basic block switching.  }
    \label{fig: compile-intro}
\end{figure}

One of the primary challenges limiting the usability of CGRA is the automation of application deployment to fully exploit its parallelism. 
In fact, deploying applications to effectively exploit parallelism requires substantial effort~\cite {revamp,8050238}.
Most existing works simplify the CGRA compilation by limiting the compilation scope to a single DFG~\cite{chordmap,Plasticine}, as highlighted in blue in Figure \ref{fig: compile-intro}.
A common approach to DFG mapping is through modulo scheduling (MS)~{\cite{MeiDATE2003,mei2002dresc,SPR}}. This technique aims to overlap loop iterations, thereby minimizing the initiation intervals ($II$) - the time gap between the start of consecutive loop iterations.

\begin{figure}[t!]
    \centering
    \includegraphics[width=0.68\linewidth]{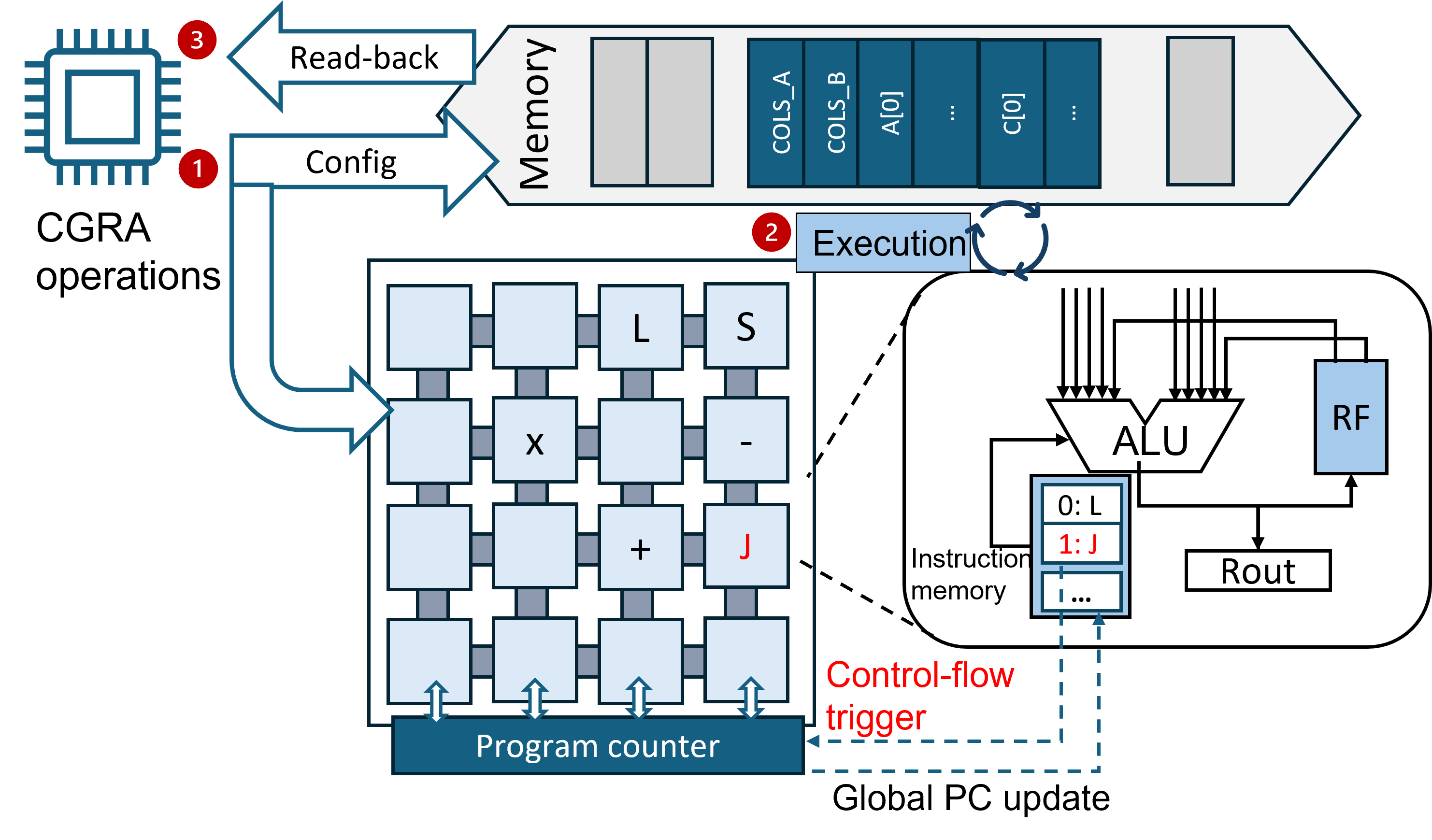}
     \caption{{
     Deployment of accelerated kernels on CGRA platforms. The CPU and CGRA must exchange inputs and outputs e.g. by employing a shared memory space. Moreover, configurations must be programmed in the CGRA accelerator prior to their execution. }}
    \label{fig: deploy}
\end{figure}

Although the DFG-based mapping strategy effectively exploits loop-level parallelism \cite{riptide,compile_conclusion}, it entirely offloads control flow management onto the host processor, as shown in Method 1 in Figure \ref{fig: compile-intro}.
Each time a kernel is deployed on the CGRA mesh, {the corresponding configurations must be loaded onto the instruction memories~\cite{10546646}}, and the input data transferred, before CGRA execution, as illustrated in Figure~\ref{fig: deploy}..
Then, upon completion of a kernel, the data must be written back into the host processor's memory space. We herein address this shortcoming by increasing the granularity of acceleration kernels, from that of DFGs to that of entire Control Flow Graphs (CFGs).

Control Data flow graph (CDFG) mapping reduces the communication overhead between basic blocks by supporting a larger scope~\cite{10.1007/978-3-030-79025-7_5,7393298}, as CDFG can span multiple basic blocks and capture control and data dependencies.
Most of the CDFG-based approaches proposed map each DFG within the CDFG independently~\cite{sgmf,4dcgra_2019}, as illustrated by Method 2 in Figure~\ref{fig: compile-intro}, relying on specialized hardware to manage the ensuing branches in the execution flow.
When a control dependency triggers the execution of a different DFG --such as from BB0 to BB1 in Figure~\ref{fig: compile-intro}—in these works (a) new instructions must be loaded for newly scheduled blocks and (b) the relevant data must be propagated accordingly~\cite{control_plane}. 
Hence, these approaches still incur run-time overheads and energy inefficiencies due to instruction and data reconfiguration~\cite{Snafu}.

Therefore, we argue that the limited support for complex control flows is a major limitation in state-of-the-art CGRA compilers, which are either limited in scope or adopt ad-hoc hardware solutions.
In contrast, our compilation method adopts a program counter (PC)-based control flow management model for time-multiplexed (also known as multi-context) CGRAs.
This allows the application to be compiled and configured once, eliminating the need for reconfiguration during basic block switching, as illustrated in Method 3 of Figure~\ref{fig: compile-intro}. 
This strategy is applicable to most CGRA architectures, which often use PC-controlled directional branches to minimize instruction memory usage~\cite{HyCUBE, openedge, morpher, miyamori1999remarc, taylor2002raw, revamp}. 
Therein, the PC-based approach is commonly limited to scheduling a single loop, where it allows for branching back to the entry point of the loop kernel--represented by the loop continuation edge in Figure~\ref{fig: compile-intro}. We instead show that such a position can be generalized by leveraging compiler transformations, in order to accommodate arbitrary control flows.

{To support CDFG mapping on a CGRA without runtime reconfiguration, we assume an abstract hardware model in which a global PC controls the execution of all PEs. Each PE is equipped with a local instruction memory. When a control operation (e.g., a branch or jump) is triggered, it sends the destination signal to the global PC, causing all PEs to synchronously switch to the target instruction address.
}
{To support data propagation across basic blocks,} our mapper leverages liveness analysis ~\cite{compiler_design} (traditionally used in register allocation to prevent assigning the same physical register to variables that live simultaneously) to preserve data consistency in any CFG, regardless of their sizes or connectivity.
For example, in Figure~\ref{fig: compile-intro} value \textcircled{2} is live-out from BB0 and live-in to BB1. 
Therefore, the mapping of value \textcircled{3} must be placed in {the neighbor of \textcircled{2} to access its produced value.}
Under the mapping constraints imposed by liveness analysis across  CFGs, basic block transitions are executed directly via branches without incurring any reconfiguration delays, as mapping live-in/live-out values is consistently performed in-between predecessor and successor blocks.

Our work considers a PC-based control model for any CDFG compilation that avoids reconfiguration overhead. 
Das et al.~\cite{pc_control_mapping} also consider a similar hardware model, in which mappings are repeatedly sampled from a search space until data consistency is achieved.
Instead, we rely on liveness analysis to draw the feasible mapping space systematically.
Moreover, as noted in \cite{pc_control_mapping}, loop-level parallelism is not supported, since a control operation must be executed at the end of each basic block to determine the subsequent execution path, thereby requiring each iteration to wait for the previous one to complete.
To overcome this issue, we adopt a CFG-based transformation to adapt a new loop kernel with a loop length $L'$ equal to $II$, as shown in Method 3 in Figure~\ref{fig: compile-intro} 
Our approach ensures that our CDFG mapper provides the same scheduling efficiency of loop DFG mapper alternatives.

In summary, in this work, we present a modular compilation framework for CGRAs\footnote{ The code base is open-sourced in https://github.com/esl-epfl/Compigra.git},  offering two key advantages over traditional state-of-the-art~(SoA) mapping methodologies: \textbf{(1) we extend the scope of the compilation to the mapping of CDFG in a PC-based hardware model, eliminating the reconfiguration overhead associated with basic block communication}; and \textbf{ (2) we improve the efficiency of runtime in the PC-based control model through CFG transformations to support modulo scheduling}.
In more detail, our main contributions are listed as follows:
\vspace{-0.5em}
    \begin{itemize}
        \item We present an end-to-end compilation framework comprising hardware compatibility transformation, efficiency optimizations, and parameterized hardware mapping to support control flow management for various applications on a {homogeneous} PC-based CGRA abstraction.
        \item We design a customized \texttt{cgra} dialect in MLIR that directly represents CDFGs at the IR level and cleanly separates hardware-agnostic from hardware-specific optimizations. 
        \item {We leverage liveness analysis to formulate an integer linear programming (ILP) model for mapping each basic block’s DFG onto the CGRA. When the ILP fails under hardware constraints, we introduce a post-mapping CDFG repair mechanism that resolves infeasibility without re-scheduling solved basic blocks, thereby significantly reducing mapping complexity.}
        
        \item {We show how conventional CFG optimizations (e.g., basic block fusion) and hardware-aware techniques (e.g., modulo scheduling) can be composed with PC-based control hardware, enabling robust CDFG mapping that remains compatible with state-of-the-art optimization tools.}

        \item We compare our methods with state-of-the-art alternatives, achieving speedups of 2.12×, 1.48×, and 1.20× over EffiMap~\cite{pc_control1}(PC-based CDFG mapping), SAT-MapIt~\cite{satjournal}(DFG mapping + CPU support), and Marionette~\cite{control_plane}(CDFG mapping with reconfigurable supported hardware on CGRA), respectively.
        

    \end{itemize}
    \vspace{-0.5em}
    
The rest of the paper is organized as follows: Section~\ref{sec: RWs} reviews related work on SoA compilation methods for CGRAs. Section~\ref{sec: motivation} outlines the motivation for a reconfiguration-free CDFG compilation framework. Section~\ref{sec: framework} presents the overall compilation framework, while Section~\ref{sec: middle-end} details hardware-agnostic transformations followed and hardware-specific IR conversion. 
Section~\ref{sec:optimization} details the CFG transformation to enable modulo scheduling. Section~\ref{sec:mapping} describes our CDFG mapping approach using an ILP model for each basic block, incorporating inter-block dependencies identified through liveness analysis. Section~\ref{ref: exp framework} outlines the experimental setup, and Section~\ref{sec:res} presents our evaluation and results.

\section{Related Works}
\label{sec: RWs}

\subsection{MLIR}
Multi-Level Intermediate Representation (MLIR)~\cite{MLIR} comprises a front-end translating high-level languages (we herein consider C/C++) into intermediate representations (IRs). 
In particular, MLIR supports dialects (i.e. IR variants), allowing compiler designers to define domain-specific operations, types, and attributes~\cite{ml_cgra,mlir_cgra}. 
As a sub-project of LLVM, MLIR ensures compatibility with commonly used LLVM optimization passes and the C/C++ parsing front end. We take advantage of the ability of MLIR to represent both the control flow and the data flow at the IR level, which facilitates graph transformations in the middle end. 
By integrating our defined hardware-specific operations into MLIR, we achieve a one-to-one operation mapping to the hardware in the back end.

\subsection{Data flow graph mapping}

\begin{figure}[t!]
    \captionsetup{type=table}\caption{Comparison with state-of-the-art CGRA mapping methods.}
    \centering
    \includegraphics[width=0.95\linewidth]{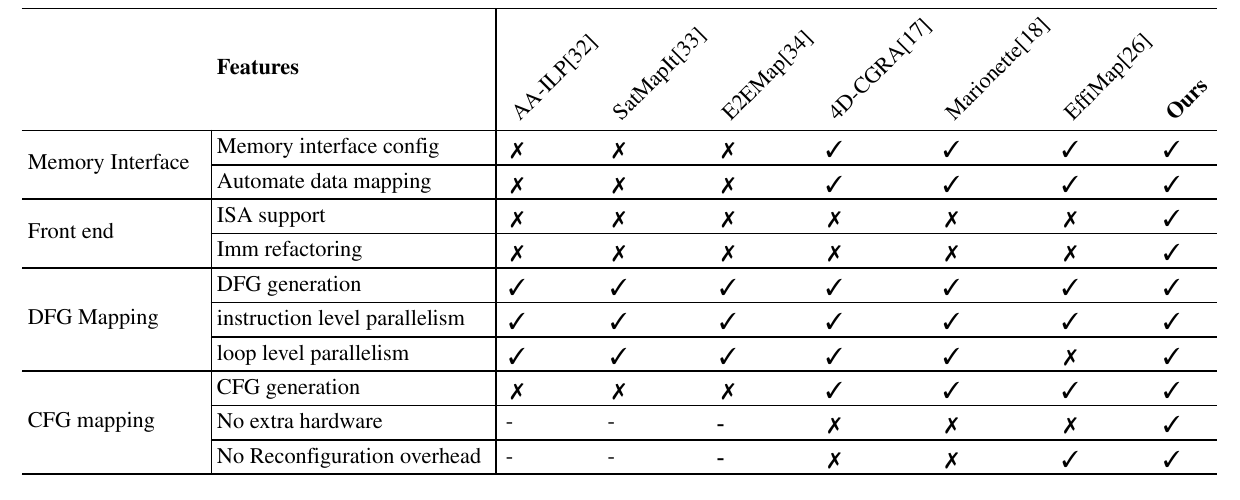}
    \label{tab: SoA comparison}
    \vspace{-2em}
\end{figure}

Table \ref{tab: SoA comparison} includes the various features supported by SoA mappers.
DFG mappers, such as AA-ILP\cite{chin_architecture-agnostic_2018}, SAT-MapIt\cite{satmapit}, and E2EMap\cite{liu_e2emap_2024} perform modulo scheduling, partially overlapping loop iterations, minimizing their Initiation Interval (\textit{II}) to reduce runtime.
Modulo scheduling handles both mapping and runtime optimization for a single loop but has limitations in managing control flow, e.g. it can only issue control-free iterations to the CGRA mesh, limiting flexibility. 
In these works, data transfer operations are not automated, and DFG mappers can process only one basic block because they do not support CFG generation, thus limiting the coverage of the CGRA code.

Recent efforts to expand such scope include the IR transformation, which merges multiple DFGs into one, employing traditional compiler optimizations such as loop unrolling and partial prediction for if conversion~\cite{7529868,10531698,10.1145/2459316.2459319}.
Similarly, Dual-Issue Single-Execution (DISE)~\cite{dynamic_ii, compile_branch2} issues two branches (if-else clauses) within a single loop to the same PE, allowing the correct execution path to be determined at runtime based on the condition. 
{
MLIR-CGRA~\cite{mlir_cgra} decouples standard loop optimizations from CGRA-specific mapping passes and is architecture-agnostic, effectively exposing spatial and temporal parallelism in structured loop kernels. However, its compilation flow fundamentally targets DFGs derived from single loops or loop nests.}
Although these methods extend the applicability of modulo scheduling, {they still cannot guarantee successful compilation of all irregular loops, such as loops with variable bounds, thereby limiting the range of applications that can be accelerated by CGRAs.}~\cite{compile_branch1}.
Indeed, in these methodologies, the compilation scope remains constrained, particularly for nested loops and divergent branches. Instead, we support general application compilation, enhancing mapping efficiency.

\begin{figure}[t!]
        \centering
        \includegraphics[width=\textwidth]{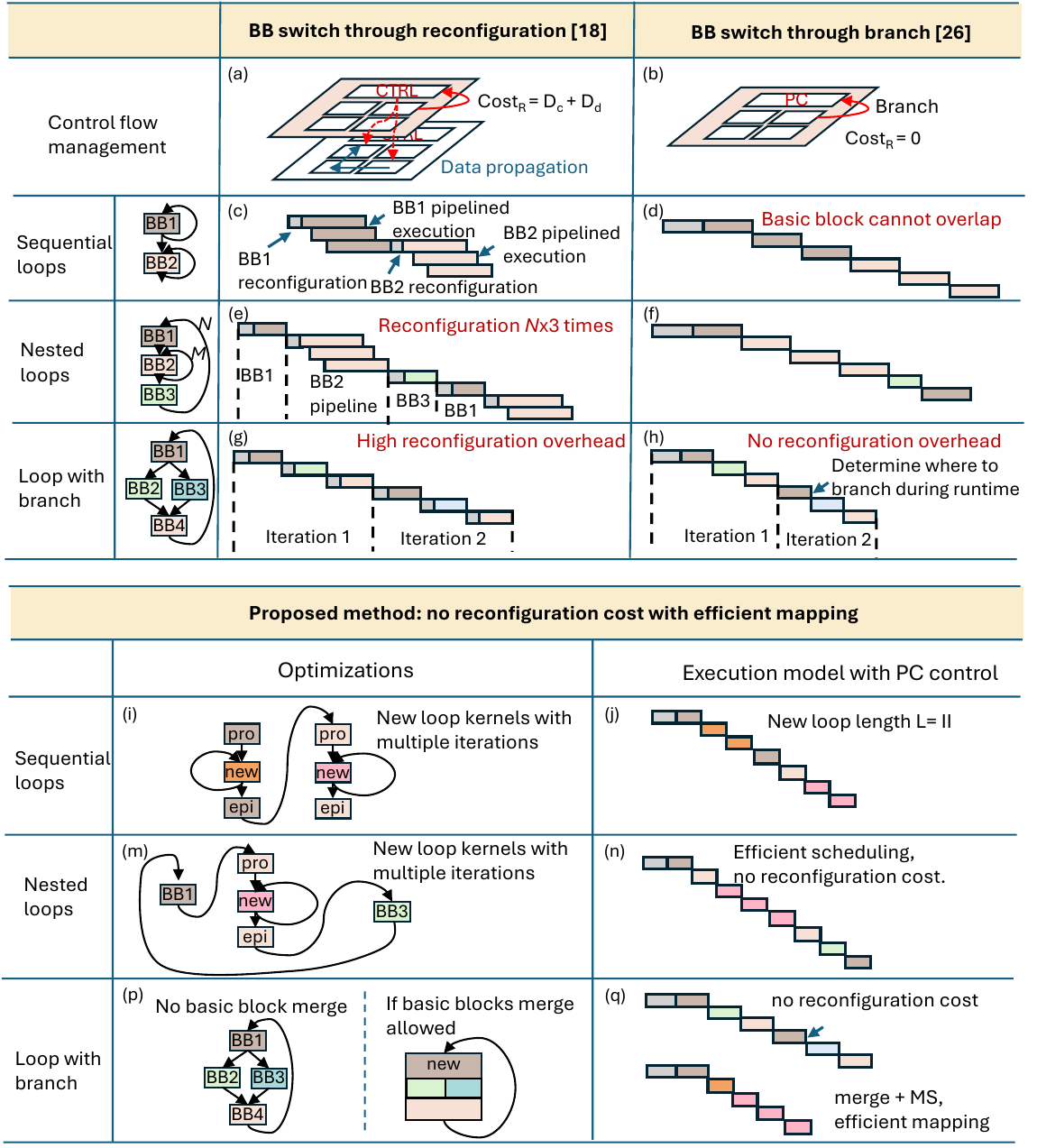}
        \caption{Reconfiguration-based CDFG mapping incurs reconfiguration overhead, while previous PC-based mapping strategies struggle with low scheduling efficiency. Our approach achieves no reconfiguration costs and efficient mapping through compiler optimizations.}
        \label{fig:execution}
        \vspace{-1em}
\end{figure}

\subsection{Control data flow graph management with reconfiguration}\label{sec:decouple2}

Compared to DFG mapping, the challenge of control management arises from compiling an entire CDFG composed of multiple basic blocks, each containing a DFG that captures computational dependencies while minimizing control overhead.  
To expand the range of applications compiled onto CGRAs, approaches that integrate ad-hoc hardware components for control management have been proposed.
In \cite{riptide,cfg_intro}, independent mapping of DFGs for each basic block is applied, while specialized hardware is incorporated for DFG switching.
Dora~\cite{Dora} introduces dynamic binary code generation at runtime, leading to compilation and reconfiguration delays. 
The 4D-CGRA~\cite{4dcgra_2019} and Marionette PEs~\cite{control_plane}  optimize the reconfiguration methods via specialized hardware design (see Table \ref{tab: SoA comparison}).
4D-CGRA~\cite{4dcgra_2019} achieves branch prediction through entry instruction matching, requiring reconfiguration for the selected block instructions and data.
The marionette PE~\cite{control_plane} minimizes control overhead by enabling autonomous control information sharing among neighboring PEs, reducing it to a single clock cycle. However, data propagation through basic blocks remains a bottleneck for reconfiguration.
{However, both prior works introduce reconfiguration overhead for basic block switching.}

Figure \ref{fig:execution}.a illustrates the execution process for control flow management with reconfiguration.
{The gray bars represent the reconfiguration overhead associated with instruction and memory reconfiguration.}
The instruction reconfiguration process introduces control overhead~\cite{app_stream,cfg_compile1}, denoted as \(D_c\). The data required for the newly configured basic block is propagated from the producer PE of the previously executed block or loaded from memory, incurring additional data movement overhead, denoted as \(D_d\). 
Since control flow and data flow are fully decoupled, DFG mapping is applied independently to each basic block.
The innermost loop execution without control overhead is pipelined through modulo scheduling, as shown in Figures \ref{fig:execution}.c and \ref{fig:execution}.e.  
For sequential loops, reconfiguration is performed only during basic block switching, making it efficient with low reconfiguration overhead. 
However, the overhead increases significantly for nested loops.  
Additionally, loops with branches face a major computational bottleneck, as each basic block must be fetched separately during runtime. This limitation is illustrated in Figure \ref{fig:execution}.g.

\subsection{Control data flow management using the CGRA program counter}
CGRA PEs with program counter{s} achieve basic block switching through branches ~\cite{miyamori1999remarc,taylor2002raw,revamp}, as illustrated in Figure \ref{fig:execution}.b.  
Wang et al.~\cite{wang2025mlir} propose an integer linear programming (ILP) model for CDFG management, aiming to minimize application runtime. However, this approach suffers from exponential complexity growth as the application size increases and does not guarantee feasible solutions for general applications.  
Focusing on this challenge, EffiMap~\cite{pc_control_mapping} in Table \ref{tab: SoA comparison} introduces a bi-directional branch in hardware to handle divergent control flows. Based on that, they introduce a heuristic scheduling method for CDFG mapping at the basic block granularity that sequentially places consumer operations to their producer PEs and stores data in internal register files. 
When a valid mapping cannot be found for a basic block due to register overflow or routing limitations, the algorithm backtracks to previously scheduled blocks until a feasible solution is discovered, significantly increasing the compilation time.  
Our proposed mapping shows that CDFG can be decomposed into individual DFG mappings through liveness, as discussed in Section~\ref{sec:Decomposition}.
Furthermore, we theoretically prove that transformations to overcome register overflow and routing limitations do not require backtracking to change established scheduling, as discussed in Section~\ref{sec:failure}. 
These two strategies save compilation efficiency by decomposing the mapping complexity and avoiding rescheduling.

Moreover, their strategy results in inefficient resource utilization without exploiting loop-level parallelism.
Figure \ref{fig:execution}.h illustrates this execution model for a loop with branching, where the PC dynamically controls the transitions between blocks based on runtime conditions.  
However, since basic block execution cannot overlap under PC-based control, existing methods suffer from low scheduling efficiency.  
Figures \ref{fig:execution}.d and \ref{fig:execution}.f illustrate loop scheduling, where iterations are executed sequentially, failing to exploit loop parallelism with a low hardware utilization ratio.

\section{Motivation}
\label{sec: motivation}

As discussed above, the existing CGRA compilation approaches either lack support for CDFG mapping or exhibit runtime inefficiencies~\cite{control_plane, riptide, pc_control_mapping}:
reconfiguration-based methods incur instruction and data reconfiguration overhead during basic block transitions, while the existing PC-based method does not exploit loop-level parallelism.
Driven by these limitations, we propose the development of an end-to-end compilation framework designed to address the following challenges:

\textbf{\textit{Challenge 1:  The compiler must handle diverse applications with low hardware complexity for control flow management.}}
A general-purpose compiler should support CDFG mapping to accommodate general control flows, including divergent branches and imperfect loops,  {while respecting practical architectural constraints, in particular the instruction memory capacity of each PE.}
State-of-the-art approaches that implement this requirement with dedicated CGRA control modules that handle reconfiguration and data propagation across BBs, incurring timing penalties and area cost ~\cite{control_plane}.
Compiler-based approaches can dramatically reduce both overheads, as they rely on simple program counter manipulations at run-time to handle branches{~\cite{pc_control1}}. 
However, such a strategy has only been considered for a simple control structure, such as branching to the entry point of a control-free loop. We show that, by applying liveness analysis to basic blocks to track values propagated across them, compiler-based compilation can handle complex CDFGs with nested loops and divergent branches.

\begin{figure}[t!]
  \centering
  \begin{subfigure}[t]{0.4\textwidth}
    \includegraphics[width=0.9\textwidth]{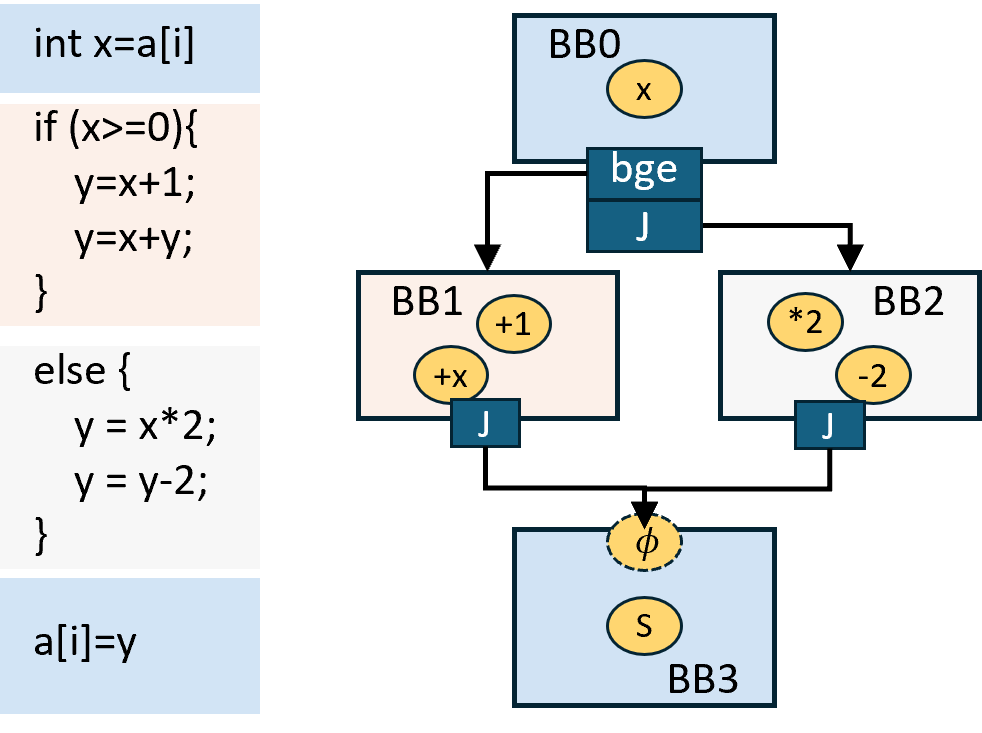}
    \caption{C source and CDFG of If-branch: BB0 determines the direction through $x$, both BB1 and BB2 jumps to BB3 after computation.}
    \label{fig:if cfg}
  \end{subfigure}
  \hfill
  \begin{subfigure}[t]{0.55\textwidth}
    \includegraphics[width=0.9\textwidth]{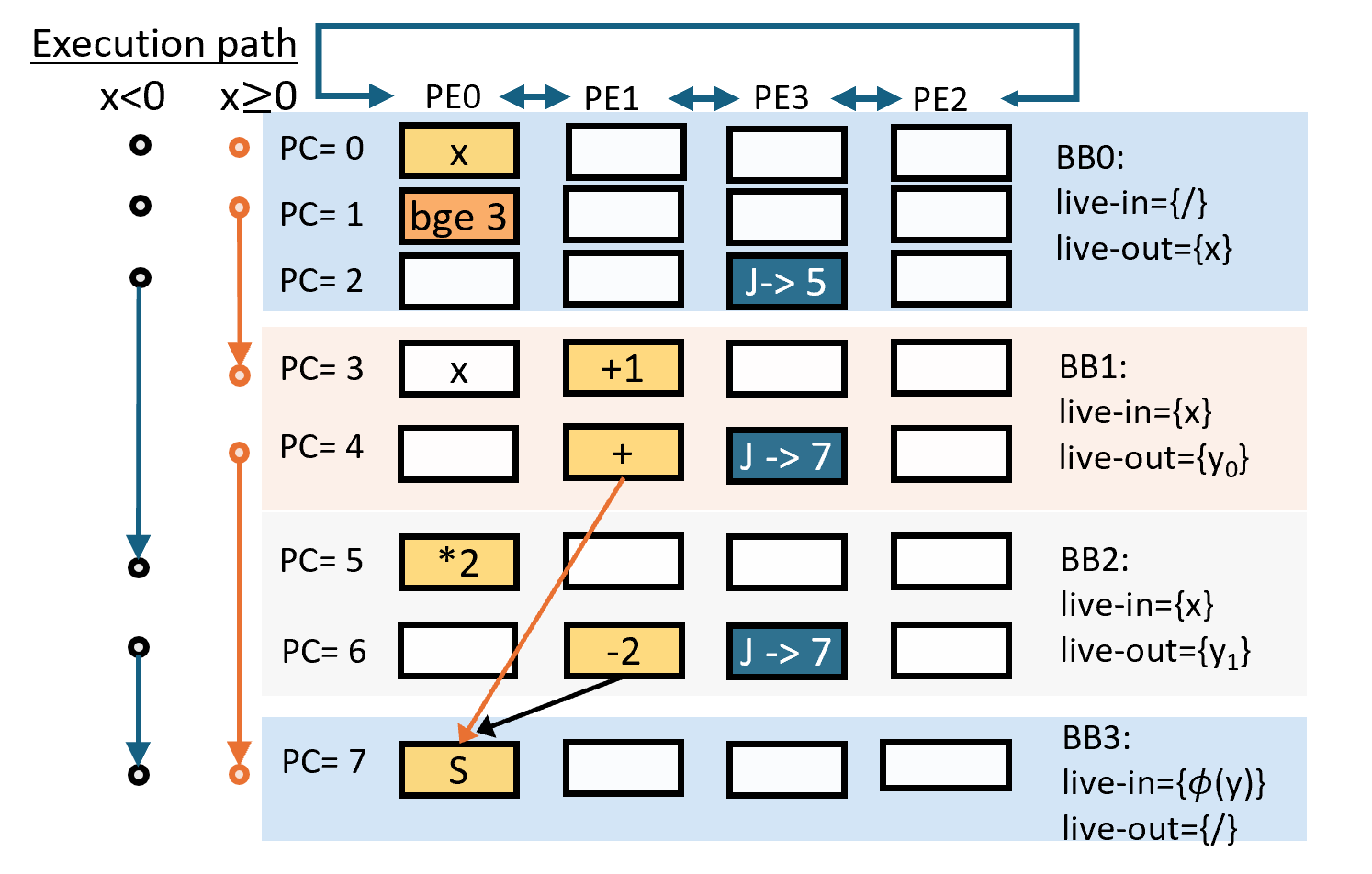}
    \caption{{If x$\geq$0, PC branches to 3 to perform +1 (Orange path). Otherwise, it jumps to PC=5 to perform the `else` clause (Blue path). PC=7 stores the result from its right neighbor regardless of which path is taken.}}
    \label{fig:if map}
  \end{subfigure}
  \hfill
  \begin{subfigure}[t]{0.41\textwidth}
    \includegraphics[width=0.9\textwidth]{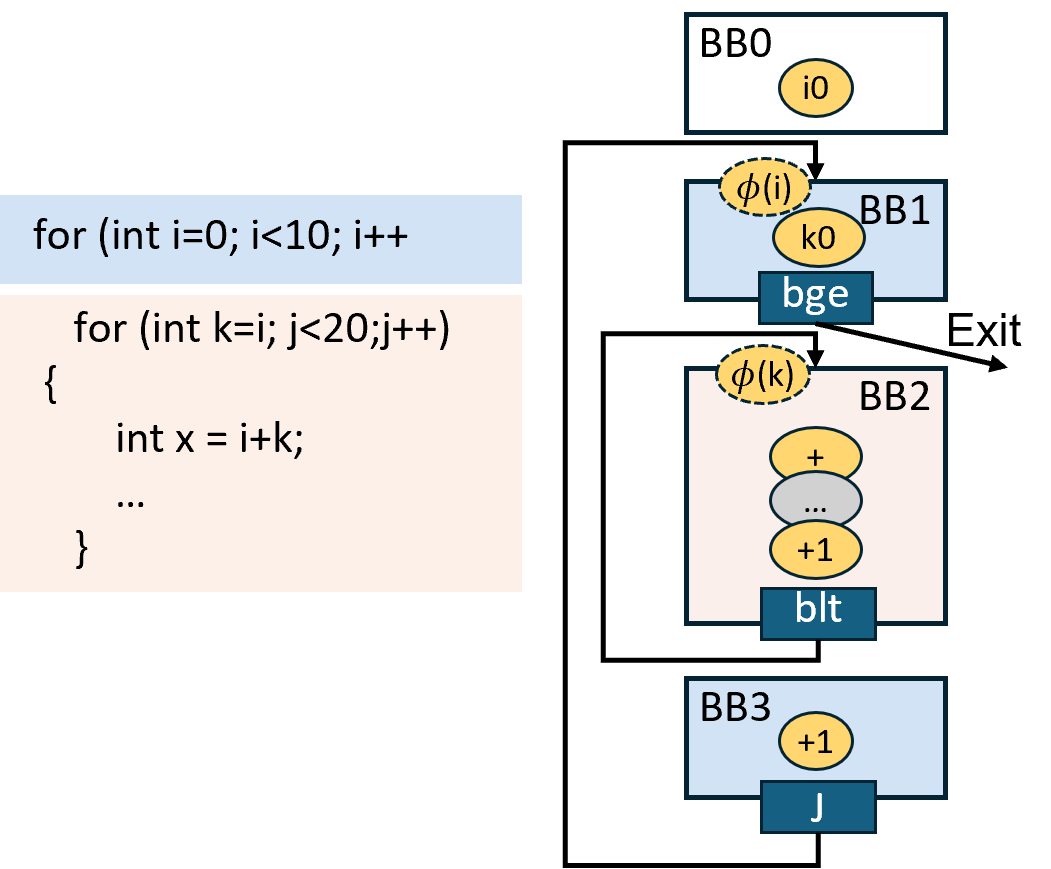}
    \caption{C source and CDFG of nested loops: BB1 quits the outer loop if $\phi(i) \geq 10$, BB2 continues the inner loop if $\phi(k)+1<20$. .}
    \label{fig:loop cfg}
  \end{subfigure}
  \hfill
  \begin{subfigure}[t]{0.55\textwidth}
    \includegraphics[width=0.95\textwidth]{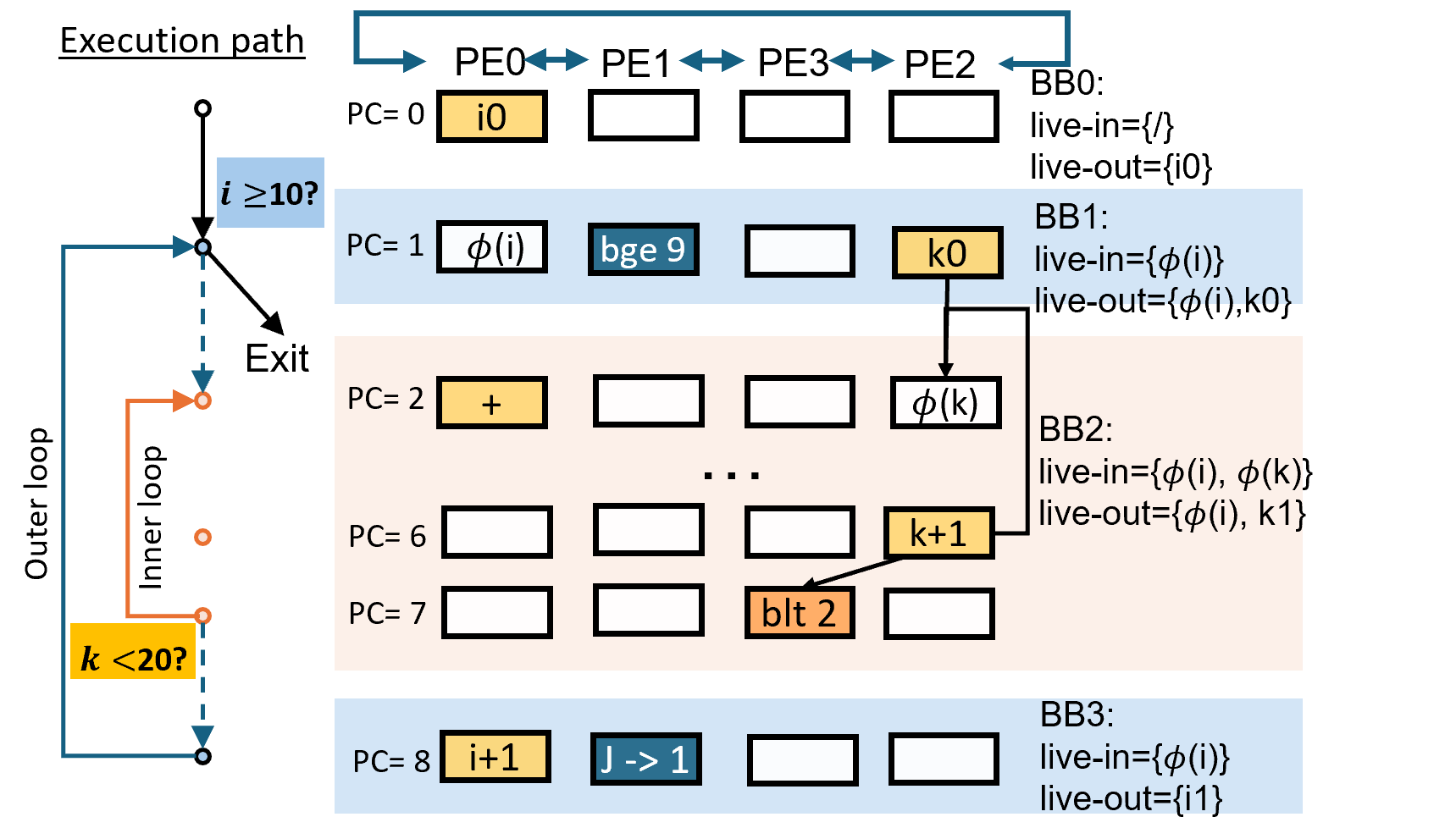}
    \caption{{In PC=1, bge determines the branch direction of the outermost loop based on the loop count $\phi(i)$, which is produced in BB0(i0) or BB3(i1) In PC=7, blt controls branches to PC=7 (continue) or PC=8 (quit).}}
    \label{fig:loop map}
  \end{subfigure}
  \caption{Mapping CDFG onto 2x2 CGRAs with liveness analysis.}
  \label{fig:why_liveness}
  \vspace{-1.5em}
\end{figure}

Figure~\ref{fig:why_liveness} presents examples of mapping CDFGs of applications on a 2x2 PC-based CGRA, {where the right side illustrates the corresponding mapping solutions. Each column represents the instructions stored in the local instruction memory of a PE, and each shaded region groups the instructions belonging to the same basic block (BB), as defined in the IR shown on the left.} 
Figure~\ref{fig:why_liveness}.a shows a code snippet with divergent branches and its corresponding CDFG.
BB0 handles the divergent branch control behavior according to the loaded value $x$.
The loaded value $x$ is the live-out of BB0 and live-in for BB1 and BB2.
To ensure correct execution, the consumer operations in BB1 and BB2 must be placed on PEs that can access the live-in value $x$—specifically, PE1 for BB1 and PE0 for BB2 in the example case.
Additionally, the final results of these computations, used to produce the value of $y$, must be mapped to the same  hardware resources to maintain consistency. 
This design eliminates the need to explicitly select the source of $y$, as represented by the $\phi$ function in BB3.
BB3 always stores the live-in value from the exact physical location (PE0 in the example), regardless of the execution path.
Figure~\ref{fig:why_liveness}.d further shows the mapping solution for a nested loop shown in Figure~\ref{fig:why_liveness}.c, where the branch operations determine the control direction during runtime.
The examples demonstrate that CDFG mapping can be decoupled into DFG-level compilation, provided inter-block dependencies are resolved using liveness analysis.
With CFG constraints respected, each DFG can be compiled individually, as intra-block values do not affect the scheduling of other blocks.

\textbf{\textit{Challenge 2: The compiler should optimize for runtime efficiency. }}
To this end, we rely on BB fusion to increase performance, since larger BBs offer more opportunities to exploit instruction-level parallelism.
Figure \ref{fig:execution}.p illustrates an example of basic block fusion, as discussed in Section \ref{sec:control optim}.
When entire loop bodies are fused in a single BBs, loop-level parallelism can be exploited to further improve runtime efficiency.
We fulfill this through CFG adaptation introduced in Section \ref{sec:optimization}.
Figures \ref{fig:execution}.i and \ref{fig:execution}.m illustrate such optimizations for pipelinable loops, where the adapted loop length \(L\) equals the initiation interval ($II$), achieving a speedup of $L$/$II$ compared to sequential execution.
The optimized execution models shown in Figure~\ref{fig:execution}.(j, n, and q) demonstrate higher runtime efficiency compared to both the models in Figure~\ref{fig:execution}.(c, e, g) and those in Figure~\ref{fig:execution}.(d, f, h).

\section{Compilation framework}
\label{sec: framework}

\begin{figure*}[b]
    \centering
    \includegraphics[width=\textwidth]{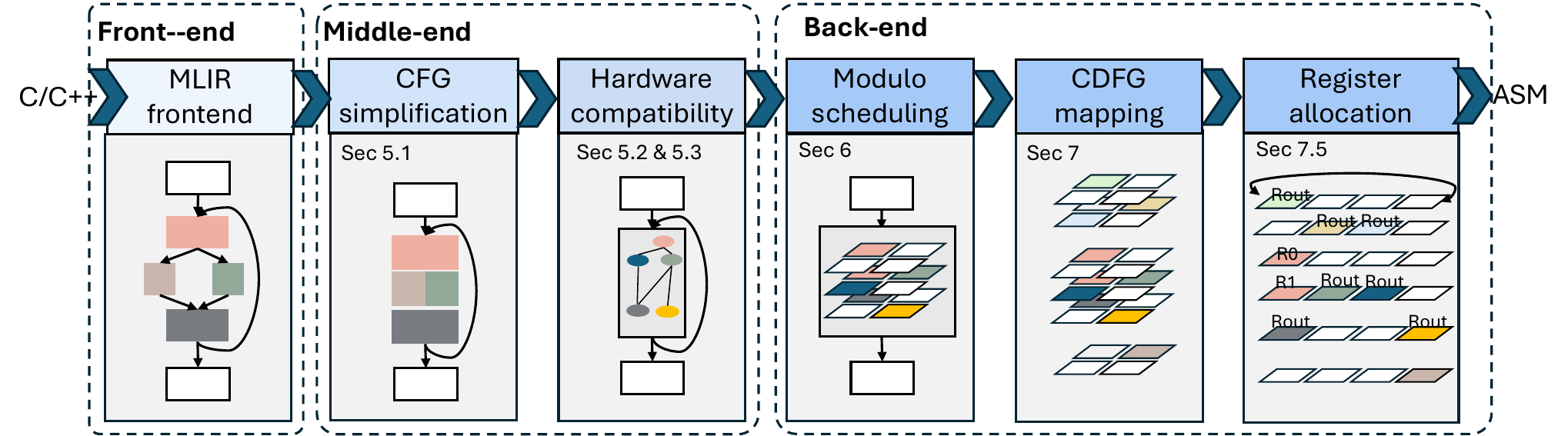}
    \caption{Birds-eye view of the proposed CDFG compilation flow,  enabling the deployment of kernel function on a CGRA meshes. The framework comprises a MLIR-based pre-processing front-end, back-end passes, ILP-based mapping,  CFG transformations to support modulo scheduling and register allocation.}
    \label{fig:framework}
\vspace{-1em}
\end{figure*} 
Figure \ref{fig:framework} illustrates a bird's eye view of the compilation process, from high-level C/C++ code to low-level assembly generation. 
We first rely on the MLIR frontend to perform code parsing and generate the CDFG abstraction at a high level, independent of any specific backend.
Section \ref{sec:control optim} introduces platform-independent optimizations, with the aim of reducing control overhead without relying on specific hardware requirements. 
Sections 5.2 and 5.3 describe hardware compatibility transformations, which are applied before any mapping-based methods to address architecture-specific differences.
As a lower-level optimization,  modulo scheduling of loop is instead discussed in Section \ref{sec:optimization}. 
The modulo scheduling results, which potentially map large subgraphs of target applications, may severely constrain the overall mapping.  
In Section \ref{sec:mapping}, we explain how we address this lack of flexibility in our framework, seamlessly integrating modulo-scheduled loops in CDFG compilation.
Finally, Section \ref{sec:RA} discusses the register allocation process, which is performed as a final step in our methodology.

\section{Middle-end}
\label{sec: middle-end}
The middle-end first performs CFG simplification to reduce the control overhead on the hardware, with the CDFG abstraction derived from the IR generated by the MLIR front end.
Furthermore, a customized \texttt{cgra} dialect is developed to handle target-specific code generation.
This design ensures compatibility between the ISA and CGRA operations, such as conditional branches and load/store instructions, while preserving a high level of abstraction. 

\subsection{Architecture-agonistic transformations}\label{sec:control optim}
To maintain a deterministic control flow, the execution of basic blocks cannot overlap. Since basic block branching dictates execution transitions, operations from different basic blocks cannot be scheduled simultaneously.  
{The transformation is architecture-agnostic and applied in the middle-end, aiming to reduce control-flow overhead.}
All transformations are independent and can be applied in any combination.
{However, their efficiency impact depends on application-specific characteristics, such as runtime-divergent branches and loop execution frequencies, as well as the backend mapping strategy and hardware implementation. In practice, these techniques are therefore applied on a case-by-case basis, and identifying an optimal combined strategy is left for future work.}

\begin{figure}
    \centering
    \includegraphics[width=1\linewidth]{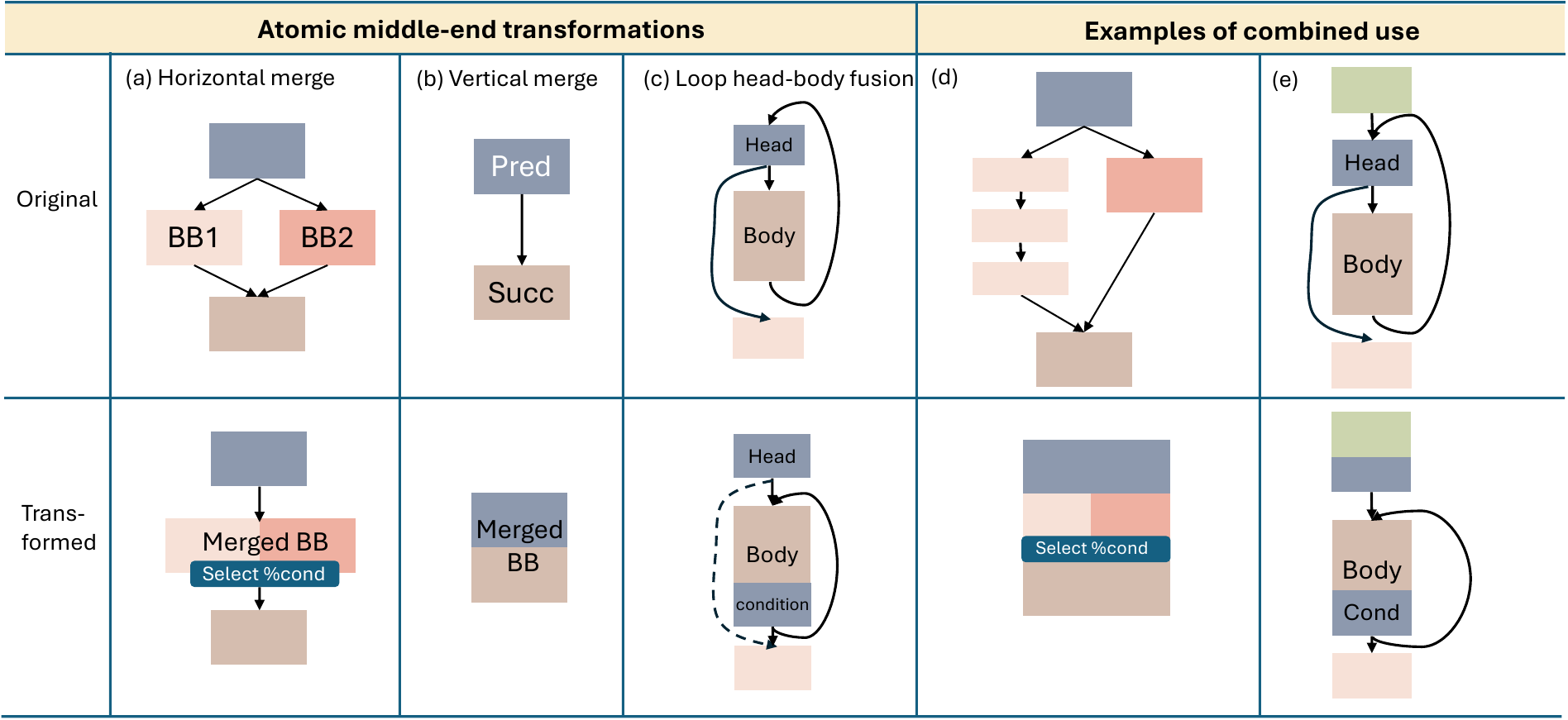}
    \caption{CFG transformations to merge basic blocks, enabling the concurrent scheduling of the merged blocks.}
    \label{fig:merge}
    \vspace{-1em}
\end{figure}

\textbf{Horizontal merge:} 
Horizontal merge {reduces control overhead through block fusion}, as shown in Figure \ref{fig:merge}.a.
Therein, a predecessor basic block has two divergent successor  blocks, that eventually converge at the same destination. 
Basic blocks can be merged if they are write-free, meaning that executing both paths only generates additional results. 
The final output is then determined by inserting a selection operation, which chooses the appropriate propagated values from the two merging blocks based on the runtime condition.

\textbf{Vertical merge:} 
Vertical merge combines two sequentially connected basic blocks into one, as shown in Figure \ref{fig:merge}.b. 
The predecessor has no alternative branches, and the successor has no other predecessors. 
Since the connection is an unconditional branch that is always taken, merging preserves the original IR functionality while allowing operations from both blocks to execute concurrently improving runtime performance.

\textbf{Loop head-body fusion:} 
MLIR front end generates the loop representation shown in Figure \ref{fig:merge}.c top.
The loop head checks the quit condition, while the loop body control flow explicitly returns to the head before continuing iteration, which disallows modulo scheduling even for control-free loop.
The transformed CFG integrates the condition check within the loop body. 
The head first verifies whether the loop should start, and then the body executes and decides whether to continue or exit based on the condition at the end of the iteration. 
Additionally, if the iteration count is known to be greater than zero at compile time, the conditional branch simplifies to an unconditional jump to the body, ensuring the first iteration always executes.

\textbf{Combining transformations:} 
Our framework adopts a modular design for each atomic transformation, providing user-defined flexibility in CFG simplification.
The CFG transformations are atomic and applied only when the topology matches specific patterns. This maintains compilation complexity while allowing flexible, iterative combinations for more complex CFG structures.  
Figures \ref{fig:merge}.d and \ref{fig:merge}.e illustrate CFG simplifications through multiple transformations. In Figure \ref{fig:merge}.d, blocks are first vertically merged, then horizontally merged with the dark pink block, and finally merged into one with their predecessor and successor. Figure \ref{fig:merge}.e shows the loop head and body fused first, followed by a vertical merge with its successor.  

\subsection{Memory operation conversion}
In the high-level abstraction of the MLIR front end, a data access is represented as a memory reference argument value. 
Our framework translates  memory references to align with the physical addresses of the corresponding data. 
To this end, on the CGRA side, memory access for arrays is decomposed into two steps: first, computing the element offset by multiplying the index with the stride; second, adding this offset to the array’s base address. For scalar values, only the base address is needed.

Such approach readily adapts to integration strategies where data transfers between CGRA and host are performed through a shared memory space, the most common strategy for CGRA system integration \cite{openedge}.
The following code snippets show the IR produced in the middle-end after memory operation conversion.
In this example, \%1 represents the offset value, and the constant 1000 serves as the base address corresponding to the argument.



\begin{figure}[h]
    \raggedright
    \includegraphics[width=0.4\textwidth]{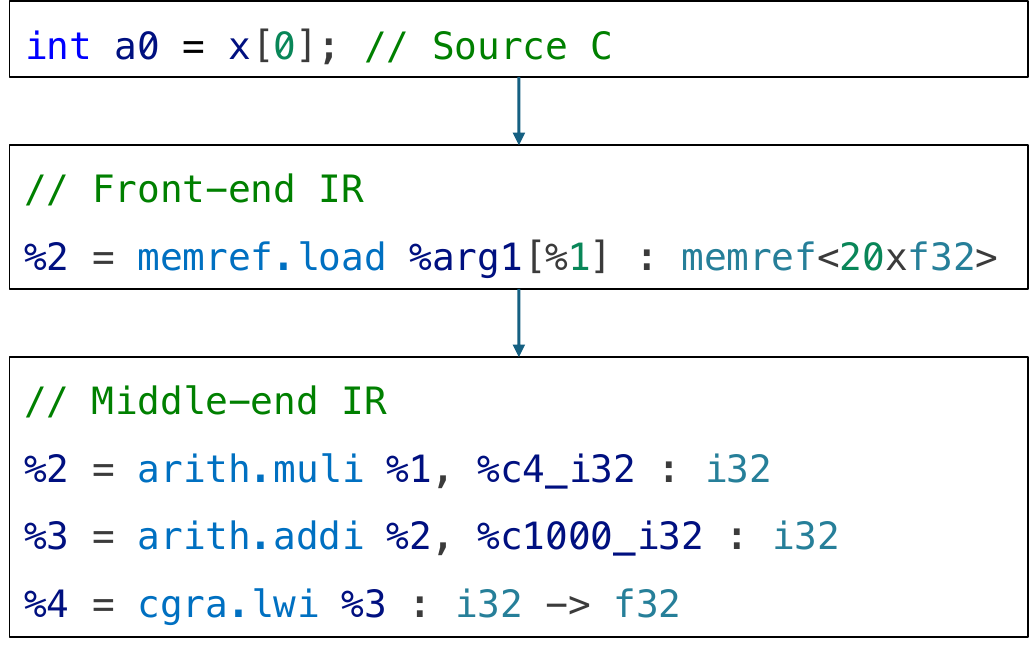}
    \vspace{-1em}
\end{figure}


\subsection{IR optimization for hardware compatibility}

Each CGRA implements its own ISA based on the supported operations and optimizations. Our compiler automatically eases the portability towards different architecture by providing a dedicated passes which
translates illegal instructions into ones supported by a target CGRA. The translation is automated by defining a platform-specific \texttt{cgra} dialect. 

For the OpenEdgeCGRA \cite{openedge} instances targeted by the experiment in Section \ref{sec:res}, the ISA translation operated by the backend rewrites the MLIR's ``compare'' and ``select'' instructions into the CGRA's corresponding instructions that use the zero and sign flags from the previous instruction. Similarly, we connect basic blocks by leveraging the CGRA's specific ``branch'' and ``jump'' instructions.

The intermediate refactoring pass also checks if immediate values encoded in instructions are compatible with ISA specifications,  never exceeding the size of immediate fields (e.g. 12 bits for additions for the ISA in Section \ref{sec:res}), decomposing them across multiple operations if required.
This middle-end transformation hence encompasses the translation of read/write IR statements to loads and stores, and the rewriting of other operations with constructs supported by the target ISA. This include transformations to generate immediate value using arithmetic operations if the hardware does not support immediate fields.

\vspace{-0.5em}
\section{IR transformation for modulo scheduling}\label{sec:optimization}
The PC-based control model requires non-overlapping execution of basic blocks, imposing two constraints that affect runtime efficiency.
First, operations from different basic blocks cannot be scheduled simultaneously. This limitation is addressed through basic block fusion, as discussed in Section~\ref{sec:control optim}.
Second, control operations such as branches and jumps must be scheduled at the end of each basic block to determine control flow direction, as illustrated in Figure~\ref{fig:why_liveness}.
This constraint affects loop blocks, which must complete the current data computation before executing the control operation. 
Thus, it hinders the exploitation of loop-level parallelism from modulo scheduling, where the next iteration starts before the current one finishes.

\begin{figure}[t]
    \centering
    \includegraphics[width=0.85\linewidth]{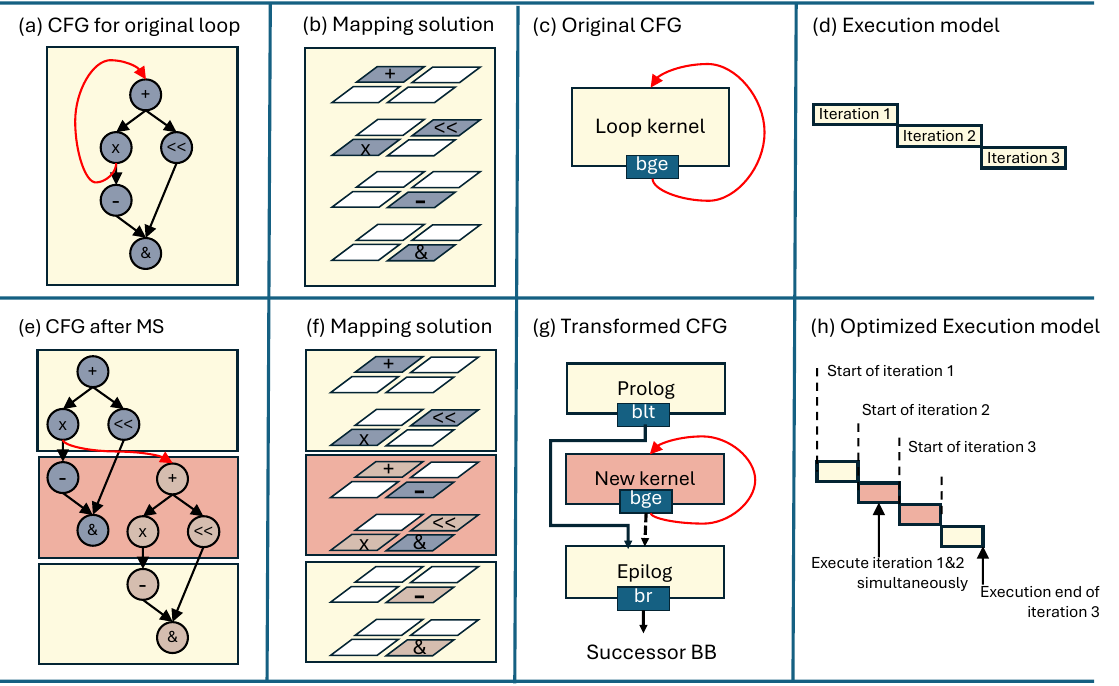}
    \caption{CFG transformation to enable loop-level parallelism on both IR level and assembly level. (a) original CFG with a single loop; (b) mapping solution for loop length $L$=4; (c) Original CFG for loop iteration; (d) original execution cycles; (e) transformed CFG with three blocks. The newly transformed loop block enables the simultaneous execution of two iterations in parallel; (f) New loop block has a reduced loop length $L'=II=2$, while the original DFG is computed in two sequential iterations; (g) New CFG graph with new loop kernel encapsulates multiple iterations; (h) efficient execution cycles. }
    \label{fig:ms-cfg}
    \vspace{-1em}
\end{figure}

To overcome this latter limitation and exploit loop-level parallelism, we describe a CFG adaptation transformation that reshapes the IR, forming a new loop kernel with an iteration length $L$ equal to $II$.
The new loop kernel includes different parts of $\lceil L/II \rceil$ iterations.
Figure \ref{fig:ms-cfg} illustrates the transformation of the CDFG with respect to a $II=2$ solution generated by a modulo scheduler. 
While modulo scheduling does not reduce the execution time of a single iteration, it improves the parallelization of loop iterations by minimizing the initiation interval.
As a result, each iteration still executes in the loop length \(L\), as illustrated by the dark blue nodes in Figure \ref{fig:ms-cfg}.f. 
However, the new loop block derived from CDFG reshaping, depicted in berry red, now encapsulates multiple iterations. This block achieves a loop length of \(L = II\), with execution remaining non-overlapping, relying on PC-based control management.

\begin{algorithm}
\caption{CFG adaptation w.r.t MS result}
\raggedright
Input: loop length $L$; initial interval $II$; 

 \texttt{// Init prologue and loop kernel}\;
\For{$l$=1 : $\lceil L/II \rceil$}{
    Init an empty basic block $B_l$.

    Create DFG in $B_l \leftarrow \mathbb{I}^l: I_k \in \mathbb{I}^l$ if $l \cdot II \leq t_k \leq (l+1) \cdot II $.

    \If{$l \ge 1$}{
    Set up the false branch flag of $B_{l-1} \leftarrow B_l$ }
}

Set up the true branch flag of $B_{\lceil L/II \rceil} \leftarrow B_{\lceil L/II \rceil}$ 

\texttt{// Create exit blocks}\;
\For{$l'$=$\lceil L/II \rceil-1 : 1 $}{
    Init an empty basic block $B_{l'}^*$;

    Set up the true branch flag of $B_l  \leftarrow  B_{l'}^* $

    Create DFG in $B_{l'}^* \leftarrow \mathbb{I}^{l'}: I_k \in \mathbb{I}^{l'}$ if $ t_k > ( l^* +1) \cdot II~\forall l^* \leq l' $.
    
    Connect $B_{l'}^*$ with the finishing block unconditional jump.

    \If{$l'=\lceil L/II \rceil-1$}{Set up the false branch flag of $B_{\lceil L/II \rceil} \leftarrow B_{l'}^*$ }
}
\label{algo: cfg_adapt}
\end{algorithm}

Algorithm \ref{algo: cfg_adapt} illustrates how CDFG-based transformations are employed to enable the integration of modulo scheduling of loops in our framework.
The transformations creates basic blocks formed by grouped operations, as shown in the rectangles in Figure \ref{fig:ms-cfg}.f. The transformed CDFG is a subgraph of the overall CDFG representing the kernel.
Each of its basic blocks contain operations executed in the $II$ time period, and a new iteration will be initiated in the next block.
Hence, a control operation dictates whether to continue or quit the loop execution. 
For example, if the current iterator dictates loop exit, the controller in the prologue (blt) in Figure \ref{fig:ms-cfg}.g branches to the epilogue and finishes the remaining operations of the current loop count.
Otherwise, the PC increments one by default and passively branches to the basic block below.
Such an approach allows us to support the execution of loops with variable, and even data-dependent, number of iterations. 
The exit blocks end up with an unconditional jump to the successor basic block of the original loop to continue the following computation.

Note that different choices for the adopted MS algorithm can be embedded in our framework. 
Indeed, we aim at mapping scope without compromising the obtained $II$, assuming a feasible schedule is provided by an existing modulo scheduling tool (we use the SATMapIt \cite{satmapit} for the results reported in Section \ref{sec:res}).
Our framework operation mapping generated by the modulo scheduler is reused, thereby reducing compilation time.
Section \ref{sec:RA} discusses how to integrate modulo scheduling outcomes into the overall compilation of arbitrary CDFGs, possibly comprising multiple loops. 

\section{CDFG mapping}\label{sec:mapping}
\begin{figure}[htbp!]
    \centering
    \begin{subfigure}[b]{0.57\textwidth}
        \centering
        \includegraphics[width=\linewidth]{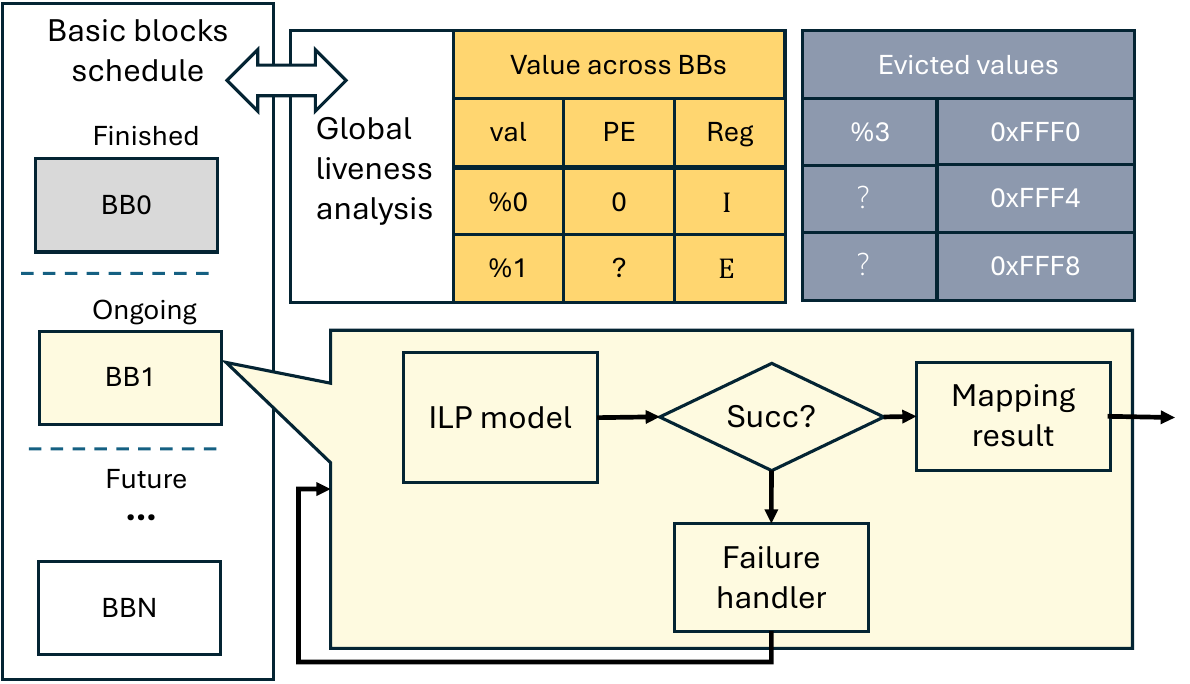}
        \caption{CDFG scheduling with individual basic block mappers.}
        \label{fig:top-scheduler}
    \end{subfigure}%
    \hfill
    \begin{subfigure}[b]{0.4\textwidth}
        \centering
        \includegraphics[width=\linewidth]{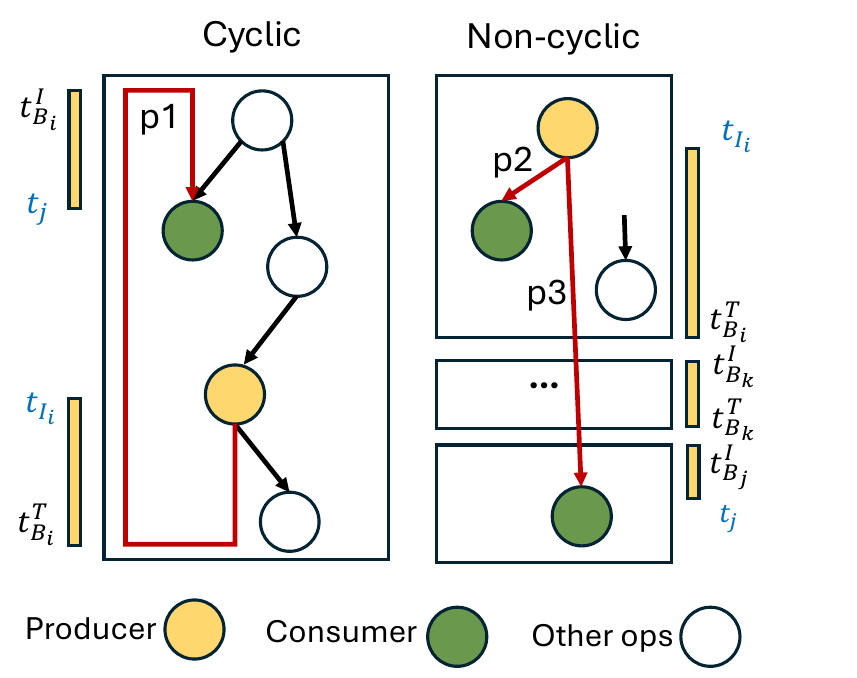}
        \caption{Computation of the execution path from producer to consumer instructions.}
        \label{fig:path-seeker}
    \end{subfigure}
    \caption{Block scheme of the scheduler framework,  coordinating basic block mappers through the placement of live-in/-out values propagated across basic blocks.}
    \label{fig:combined}
    \vspace{-1.5em}
\end{figure}

Figure~\ref {fig:top-scheduler} illustrates the scheduling methodology for mapping a CDFG, which is decomposed into DFG scheduling through liveness analysis, as detailed in Section~\ref {sec:Decomposition}. Inter-BB live values are stored in physical registers, with strategies for managing them discussed in Section~\ref {sec:in or out}. Each DFG is mapped using the ILP model described in Section~\ref {sec:mapper}. When the ILP model successfully generates a solution, the mapping results for values propagated across BBs are recorded for CDFG-level integration (highlighted in the yellow table in Figure~\ref {fig:top-scheduler}). Subsequent DFGs that consume these values must adhere to the established mappings.

If the ILP model fails to find a feasible solution, a failure handler transforms the DFG by adding routing and DFG splitting. Details of this process are provided in Section~\ref {sec:failure}. If the DFG is split in a node that is evicted to memory, the evicted values and their corresponding storage locations are recorded (highlighted in the grey table in Figure~\ref {fig:top-scheduler}), and any future references to these values are replaced by load operations from the specified memory locations. Finally, Section~\ref {sec:RA} describes the integration of all basic block mappings into the generation of the final assembly code.

\subsection{Decomposition of CDFG to individual DFGs}\label{sec:Decomposition}
A CDFG consists of connected basic blocks, allowing values to propagate through them. The interface between basic blocks and the CDFG is managed through live-in and live-out values, which define how data flows between blocks.
A value is considered as live-in for a basic block \( b \) if it is used before being defined within the block or originates from a predecessor. This relationship is expressed as:  
\begin{align}
    \vspace{-0.3em}
    In\left[b\right] &= use\left[b\right] \cup (out\left[b\right] - def\left[b\right])
    \vspace{-0.3em}
    \label{eq:live-in}
\end{align}
where \( use[b] \) represents values used within the block before being defined, \( def[b] \) includes values defined within the block, and \( out[b] \) consists of live-out values that remain live after the block.
The live-out set is computed as:  
\begin{align}
    \vspace{-0.3em}
    Out[b] &= \bigcup_{s \in Succ\left[b\right]} In\left[s\right]
    \vspace{-0.3em}
    \label{eq:live-out}
\end{align}
where \( Succ[b] \) denotes the set of successor blocks. This equation ensures that the live-out values of block \( b \) are determined by the live-in values of its successors.

These equations govern the propagation of values across basic blocks, ensuring that data dependencies are respected by tracking live-in and live-out values.
Algorithm \ref{algo:liveness analysis} shows this process, which involves iteratively calculating which variables are live at different program points by analyzing the CFG and variable usage patterns~\cite{compiler_design}.

The computed live values indicate that they are propagated across basic blocks.
Their producer PEs are initially marked as unknown, while their register index is indicated either as `External or `Internal', according to the heuristic described in the following section (see Figure \ref{fig:top-scheduler}). `E' values are stored in the output register (Rout) of PEs, allowing them to be read by neighboring cells but preventing the PE from executing other operations until the value has been completely consumed. `I' values are stored in a PE register file (RF), which does not block the computation, but makes the value only accessible locally to a PE, as shown in Figure \ref{fig: deploy}.
Once the basic block mapper (described in Section \ref{sec:mapper}) determines the PE production from their defining block, the PE index in the table is updated to ensure that the consumer(s) operation of each value can access them.

\subsection{Value assignment across basic blocks}\label{sec:in or out}

\begin{figure}[t!]
    \begin{minipage}{0.45\textwidth}
        \begin{algorithm}[H]
        \raggedright
        \small 
        \caption{Global liveness Analysis}
        \SetKwInOut{Input}{Input}\SetKwInOut{Output}{Output}
        Initialize $In\left[b\right]$ and $Out\left[b\right]$ to empty sets for all blocks\;
        
        \Repeat{{$\neg changed$}}{
            $changed \gets false$\;
            
            \For{block $b$ in reverse order of CFG}{
                $NewOut \gets \cup_{s \in Succ\left[b\right]} In[s]$\;
                
                \If{$NewOut \neq Out\left[b\right]$}{
                    $Out\left[b\right] \gets NewOut$\; 
                    $changed \gets true$\;
                }    
                $NewIn \gets use\left[b\right] \cup (Out\left[b\right] - def\left[b\right])$\;
                
                \If{$NewIn \neq In\left[b\right]$}{
                    $In\left[b\right] \gets NewIn$\; 
                    $changed \gets true$\;
                }
            }
        }
        \Return $In$ and $Out$\;
        \label{algo:liveness analysis}
        \end{algorithm}
    \end{minipage}
    \hfill
    \begin{minipage}{0.45\textwidth}
        \centering
        \begin{algorithm}[H]
            \raggedright
            \small 
            \caption{Shortest path for value \\ propagation}
            \SetKwInOut{Input}{Input}\SetKwInOut{Output}{Output}
            \textbf{Input:} Instruction $I_i$, $I_j$ and their belong blocks $B_i$, $B_j$
            
            Init $L=0$;
            
            \If{$B_i == B_j$}{
                \If{$I_i \rightarrow I_j$ is \text{back edge}}{
                    $L \leftarrow (t_{B_i}^T - t_{I_j}) + (t_{I_i} - t_{B_i}^I)$\;
                }
                \Else{$L \leftarrow$ 1\;}
            }
            \Else{
                 $\xi_{I_i \rightarrow I_j} \leftarrow \{b_m, \dots, b_n\}$, where $m,n$ is the block index of DFS CFG from $B_i$ to $B_j$\;
                 $L \leftarrow (t_{B_i}^T - t_{I_i}) + 
                ( t_{B_m}^T - t_{B_m}^I) + \dots + (t_{I_j} - t_{B_j}^I)$\;
            }
            \Return $L$;
            \label{algo: path seeker}
        \end{algorithm}
    \end{minipage}
    \vspace{-1em}
\end{figure}

The decision to store a live value in external or internal registers is based on the length of the execution path from the producer to the consumer. Figure 8b illustrates the paths from the producer $I_i$ to its consumers $I_j$, where the yellow bars indicate the liveness time frame for the producer.
The propagation path p1 in Figure~\ref{fig:path-seeker} shows the live path when the producer and consumer are located in the same (loop) basic blockand form a cyclic path. The consumer would consume the value in the next iteration. From the production to the end of the basic block, and from the beginning of the next iteration to the consumer, the producer value must be kept alive, as illustrated by the yellow bars.
If the producer and consumer reside in the same basic block and the value is consumed directly, as illustrated in p2, the live path length is 1; i.e., the consumer can consume the value in the subsequent clock cycle.
If the value is propagated across multiple basic blocks, as shown in p3 in Figure~\ref{fig:path-seeker}, its live path is determined from the producer's execution time $t_{I_i}$ to the end of its basic block $t_{B_i}^T$, through all intermediate basic blocks that must carry the value, and finally to the consumer's execution time $t_{I_j} - t_{B_j}^I$. Algorithm \ref{algo: path seeker} describes the computation of the shortest path $L$. A value produced by $I_i$ is determined to be stored in an external register if:
\begin{equation}
    \vspace{-0.3em}
    \max{L_{i->j}} \leq \theta
\end{equation}
where $\theta$ is a user-defined threshold, and $j$ is the index of its consumer. This strategy prevents a PE from being blocked by long dependency paths while ensuring the value remains accessible to all connected PEs.

\vspace{-1em}
\subsection{Basic block mapper}\label{sec:mapper}
In this section, we formulate an ILP model that aims to minimize the basic block execution time through its objective function while ensuring feasibility through a series of constraints. The constraints mainly regulate the mapping behavior through four aspects: (1) accessibility - consumers must be placed where they can access the producer; (2) dominance - consumers are scheduled before the producer; (3) liveness – a value must remain in a register until it is fully consumed; and (4) sequential memory access – the memory access sequence must be preserved.
Notably, cyclic execution patterns (e.g., loop-carried dependencies) are handled implicitly within this formulation. Such dependencies are captured through the liveness constraints and live-in/live-out analysis, which ensure correct value propagation across iterations.
The ensuing constraints are described in detail in the following.

\textbf{Global live-in constraints:}
Consumer operations must adhere to specific spatial restrictions to utilize live-ins generated outside the blocks, which are placed in internal or external registers.
If a livein variable $v_k^{\text{E}}$ is located in PE $p$ externally, then the execution units for all consumer $p_{I_j}$ where $I_j \in Cons(v_k)$ should be the neighbor or $p$ itself.
If the CGRA present nearest-neighbor connections,  the constraint is formulated as:
\begin{equation}
    \begin{aligned}
            p_{I_j} = & \text{T}(p) \ || \text{B}(p) \ ||
         \text{L}(p) \ ||  \text{R}(p\ )|| p ~ ,\forall I_j \in Cons(v_k^{\text{E}})
    \end{aligned}
\vspace{-0.3em}
\label{eq:start}
\end{equation}
where T, B, L, and R represent the top, bottom, left, and right directions that are connected to PE $p$. Additionally, before the complete consumption of the live-in value, the PE $p$ is blocked. 
\begin{equation}
    |p_{I_i} - p| \geq 0,~ \forall I_i ~ \text{if}~ t_{I_i} \leq \max{(t_j)},~ \forall t_j \in Cons(v_k^E)
\vspace{-0.3em}
\end{equation}

Otherwise, if the value is stored in a PE's internal registers, it can only be accessed by the executed PE itself:
\vspace{-0.3em}
\begin{equation}
    \begin{aligned}
            p_{I_j} = p ~ ,\forall I_j \in Cons(v_k^{\text{I}})
    \end{aligned}
\vspace{-0.3em}
\label{eq:livein-3}
\end{equation}
where $v_k^{\text{I}}$ is the internal live-in value.

\textbf{Dominance constraint:}
The dominance constraint regulates the execution order of the operations.
The execution time of  $t_{I_m}$ an operation $I_m$ should be before its consumer in the same basic block, notated as $ t_{Cons(I_m)}$:
\begin{equation}
    t_{I_m} \leq t_{Cons(I_m)}-1
    \vspace{-0.3em}
\end{equation}

\textbf{Access constraint:}
Operations should be assigned to physical PEs. In a CGRA configuration with $R$ rows and $C$ columns, the placement $p_{I_m}$ of operation $I_m$ cannot exceed the CGRA size:
\begin{equation}
    0 \leq p_{I_m} \leq RC-1
    \vspace{-0.3em}
\end{equation}

The consumer should access the producer's result, which is constrained by the hardware connectivity.
If the CGRA is grid-connected, $p_{I_m}$ has to be the neighboring PE of the producer units $p_{I_n}$:
\begin{equation}
    p_{I_m} = \text{T}(p_{I_n}) \ || \text{B}(p_{I_n}) \ || \text{L}(p_{I_n}) \ ||  \text{R}(p_{I_n}\ )|| p_{I_n} ~ ,\forall j \in Prod(I_n)
\vspace{-0.3em}
\end{equation}

\textbf{Liveness constraint:}
An operation executed afterward in the same PE would overwrite the pre-produced value. This constraint ensures the liveness of the produced results of operation $I_i$ before its consumption by $I_j$:
\begin{equation}
     | p_{I_k} - p_{I_i}| > 0 ~\forall I_k \in {\mathbb{I}}, \text{if}~  t_{I_i} 
     \leq t_{I_k} < t_{I_j}
\end{equation}
where $\mathbb{I}$ is the full operation sets.
This constraint block PEs' use until it get consumed inside one basic block.

\textbf{Control constraint:}
An operation must be executed after the initiation time $t^I$ of the basic block it belongs to, and before the control operation at the end of the block:
\begin{equation}
\vspace{-0.3em}
t^I \leq t_{I_m} \leq t_{\text{control}}, ~\forall m 
\vspace{-0.3em}
\end{equation}

\textbf{Global live-out constraints:}
The basic block must keep its live-out value inside the registers.
If the value is externally live-out, it should not be overwritten:
\begin{equation}
    \left |p_{I_i} - p \right |>0,~ \forall I_i ~ \text{if}~ t_{I_i} \geq t_j
\label{eq:liveout1}
\end{equation}
where $p$ is the PE where the live-out exists, and $t_j$ is its produce time.

For the live-out value placed in internal registers, the total number cannot exceed the available register numbers $N_r$:
\begin{equation}
    \sum_{I_i \in \text{liveout}} (p_{I_i}=p) \leq N_r, \quad \forall p
\label{eq:liveout2}
\end{equation}
where $p_{I_i}$ is the scheduling result for an operation $I_i$ which produces a live-out value.

\textbf{Memory consistency constraints:}
To maintain memory consistency, a store operation $I^S$ acts as a barrier that enforces ordering between load operations executed before and after it. Specifically, all load operations preceding the store must be completed before the store operation, and the store operation must be completed before any subsequent load operations. This is expressed as:
\begin{equation}
    t_{I_m^L} < t_{I^S} <  t_{I_n^L},~ I_m \in \mathbb{I}^L_\text{before} ~\&~ I_n \in \mathbb{I}^L_\text{after}
    \label{eq:end}
\end{equation}
where $\mathbb{I}^L_\text{before}$/ $\mathbb{I}^L_\text{after}$ represents load operation sets executed before/after $I^S$.

\textbf{Objective function:}
The final objective the ILP model is to minimize the basic block execution time,
which is 
\begin{equation}
    \min~t_{\text{control}} - t^I
\end{equation}

The objective of the optimization is to maximize runtime efficiency for the execution of the basic block. 
The constraints, defined in Equations \ref{eq:start}–\ref{eq:end}, govern the placement of operations to ensure correct execution semantics. 
Based on these constraints, the ILP model generates both the temporal and spatial mappings for each operation within the basic block.
If the scheduled operations produced value across BBs, the placement is updated in the live value table, as shown in the example of Figure \ref{fig:top-scheduler}.

\vspace{-1em}
\subsection{IR transformation to handle ILP failure}\label{sec:failure}
\begin{figure}
    \centering
   \begin{subfigure}[b]{0.6\textwidth}
       \includegraphics[width=0.8\textwidth]{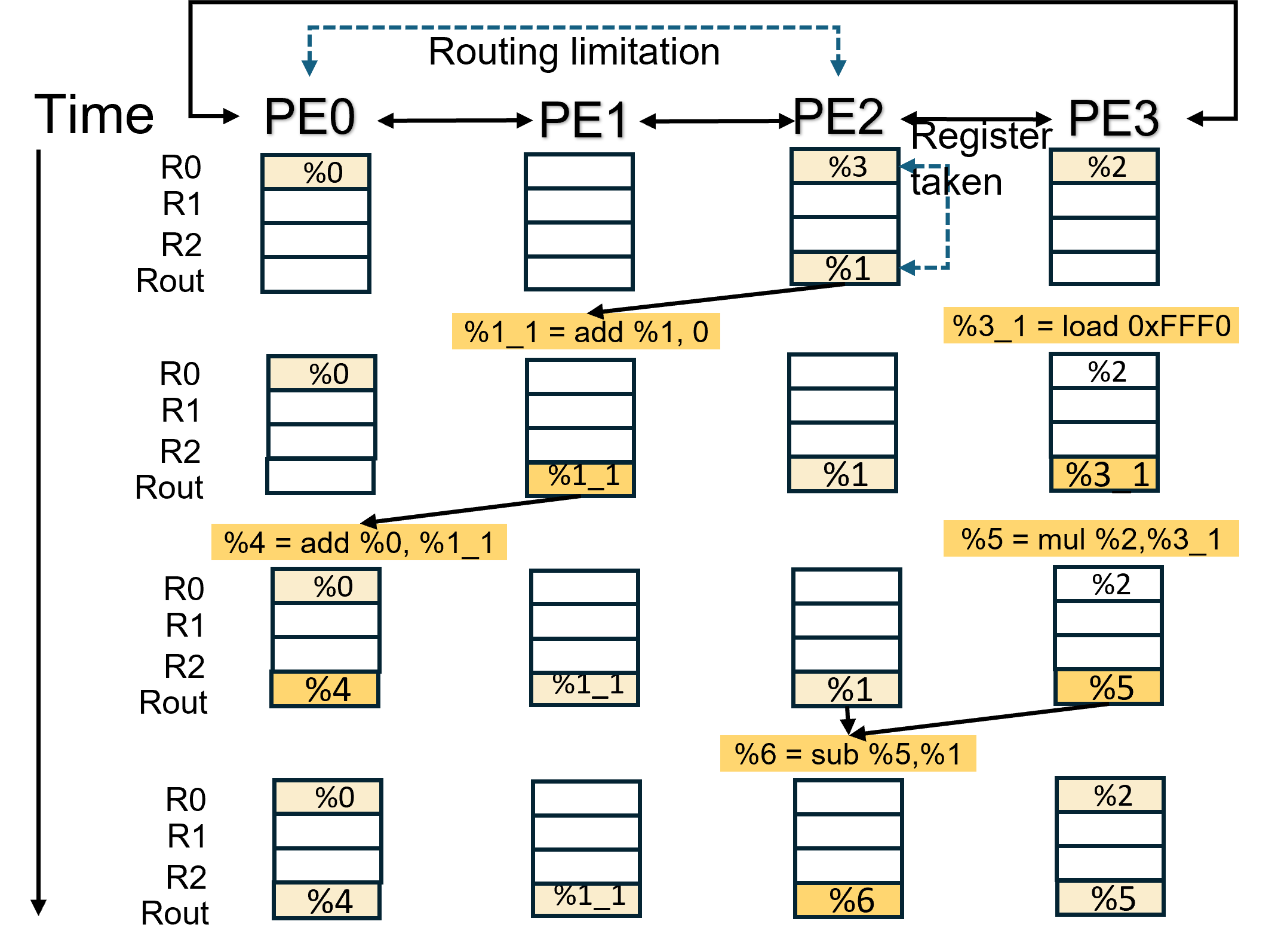}
       \caption{{Register mapping along temporal and spatial dimensions for values.}}
       \label{fig:failure map}
   \end{subfigure}
   \begin{subfigure}[b]{0.16\textwidth}
       \includegraphics[width=\textwidth]{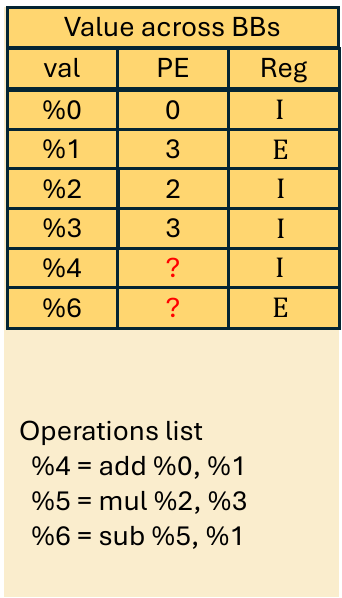}
       \caption{Pre-scheduling.}
       \label{fig:fail_model}
   \end{subfigure}
   \hfill
   \begin{subfigure}[b]{0.16\textwidth}
       \includegraphics[width=\textwidth]{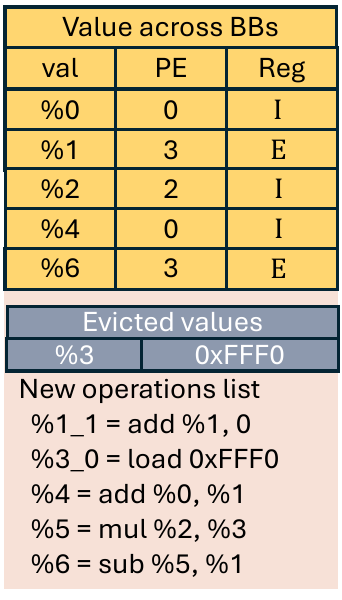}
       \caption{Post-scheduling.}
       \label{fig:suc_model}
   \end{subfigure}
    \caption{DFG transformation for converting an ILP model to find a feasible solution. The transformation takes into account limited routing and register resources (a) to schedule additional stores/loads and move operations between registers and memory (b-c).}
   \vspace{-1.5em}
   \label{fig:failure}
\end{figure}

The ILP  model does not guarantee the existence of a solution due to CGRA  hardware constraints. We herein discuss a strategy that counters two key sources of unfeasible solutions: routing limitations and register overflow.
Figure \ref{fig:failure} illustrates the IR of a block facing both failure scenarios, {where Figure \ref{fig:failure map} illustrates the spatial register storage and Figure \ref{fig:fail_model} shows the pre-scheduling operations with live value placements. }  
In the first stage, Value \%4 encounters routing limitations as \%0 is only accessible by PE0, while PE0 cannot reach PE3 for \%1. This conflict is resolved by inserting a move operation (add \%1, 0) to route \%1 from PE3 to PE1.  
Value \%5 cannot use \%2 and \%3, as they are stored in different PEs' internal registers. PE3 is blocked from popping \%3 until its external register value is consumed. To address this, \%3 is evicted to memory, and a load operation is inserted.  
Figure \ref{fig:suc_model} shows the transformed IR with additional move and load operations, allowing a feasible mapping solution.

Routing limitations arise when connectivity constraints prevent consumers from accessing values produced and stored in specific PEs.  
Register overflow occurs when internal or external registers conflict with live values, requiring one to be evicted to memory.  
While splitting the DFG by inserting load-store pairs can always resolve these issues, it may introduce significant overhead.  
To address ILP model failures efficiently, we employ a \textit{max-try or roll-back} strategy.  
If a consumer cannot access a producer, up to $N$ move operations are inserted. If a feasible solution is still not found after $N$ moves, the IR is rolled back, and a load-store pair is inserted to resolve the conflict.
The threshold $N$ depends on the load and store overhead, with $N \leq D - 1$, where $D$ represents the length of the longest routing path in the CGRA architecture.  
For example, in a 2×2 CGRA, routing from PE0 to PE3 results in $D = 2$.

Handling failures with additional move operations or load-store pairs does not require re-scheduling blocks that already have solutions.
Indeed, a basic block transformation can only affect other basic blocks by altering the result of liveness analysis, which involves the value propagated through basic blocks, as discussed below.

If a value $v$ leads to the failure of the ILP model is used and defined only inside the basic block $i$, $ v \notin use[b_j]~\forall j$, the result $ v \notin In\left[b_j\right]~\forall j$.
Moreover, $v \notin In\left[b_i\right]$ because $v \notin use\left[b\right]-def\left[b\right]$ according to Equation \ref{eq:live-in}.
In this case, the value is not recorded in the live value across BBs and cannot affect the other basic blocks.

$v$ may instead be defined and used in different blocks.
For a block $b$, if move operation $I_m$ are inserted to route the value $v$ from its original producer $I_p$, result of $I_m$ replaces the use of $v$. 
However, the set $use\left[b\right]$ still includes $v$ through $I_m$.
Therefore, the live-in set defined in Equation \ref{eq:live-in} remains unchanged, preserving the results from liveness analysis and ensuring no impact on other blocks' scheduling.

In case a load and store are inserted for  $v$, let's define  its consumer as $b_j$ and its producer block $b_i$.
For the producer block, the define and use sets remain the same: $use\{b_i\}' =  use\{b_i\}$, and $
    def\{b_i\}' =  def\{b_i\}$.
For the consumer block,  $use\{b_j\}' =  use\{b_i\} - v$, and $def\{b_i\}' =  def\{b_i\} + v_0$, where $v_0$ is the loaded value to replace $v$.
This transformation updates the value across BBs by removing $v$ from BB propagations according to Equation \ref{eq:live-in}.
Figure \ref{fig:suc_model} shows evicting value \% 3 from the register to the memory. 
This still abides by the scheduler results, as it represents a relaxation of the constraints in Section \ref{sec:mapper}, which means that the previous scheduling remains valid except that a store operation need to be inserted after the production of the split value \%3.
This process involves identifying an available neighboring PE after the split value has been produced. If no available PE exists, additional cycles are introduced to accommodate the store operation, without the need for re-scheduling.

\subsection{Composing modulo and CDFG scheduling}\label{sec:RA}
When modulo scheduling of loops is performed before CDFG transformations, the allocated PEs and registers for the live-in and live-out values of the scheduled block are first recorded as in the table in Figure \ref{fig:top-scheduler}. This ensures that subsequent blocks maintain consistency with the modulo scheduling results, so that there is no need to run the BB scheduler for the mapped blocks, as they have already been resolved.

After the scheduling for all blocks is completed, the top-level scheduler merges the mapping results of all basic blocks to obtain the CDFG mapping solution. In particular, the top-level scheduler assigned starting times ($t^I$ in Section \ref{sec:mapper}) to each basic block.
The concatenated scheduling result ensures the data consistency through the global live-in constraints (constraints \ref{eq:start}- \ref{eq:livein-3}, again in Section \ref{sec:mapper}) and global live-out constraints (constraints \ref{eq:liveout1} - \ref{eq:liveout2}).

\subsection{Back end}
As a final step for CDFG mapping, register allocation is  performed.
Operands from neighbor PEs are retrieved from their output registers, eliminating the need for register allocation in local RF when data is routed from one PE to another.
Internal register allocation is performed per-PE, similarly to state-of-the-art approaches such as~\cite{hack2006register}. 
This step involves constructing a liveness graph for each PE and applying graph coloring to assign distinct physical registers to values that interfere with one another.
After register allocation, a final rewriting pass converts the MLIR code to the input needed for the CGRA assembler. During the MLIR-to-CGRA assembler conversion, redundant jumps to basic blocks placed immediately afterwards in the instruction placements are removed.  The assembler is then translated into binary form for execution on the target hardware. 

\vspace{-.5em}
\section{Experimental framework}
\label{ref: exp framework}

\textbf{~~~Target CGRAs.} Across experiments, we considered as architectural  instances of the OpenEdge CGRA template~\cite{openedge} of varying mesh sizes.
The CGRA adopts a minimal yet Turing-complete ISA, which allows for flexible and efficient execution of a wide range of computational tasks while maintaining low hardware complexity.
The control flow is managed by an unconditional jump and four conditional branches with predicates equal (\texttt{EQ}), not equal (\texttt{NE}), greater or equal than (\texttt{GE}), and less than (\texttt{LT}).
The hardware is synthesized {at} 250MHz {in} 65nm technology and a cycle-level software simulator is provided to verify the functional correctness and evaluate the runtime metrics.

\begin{table}[htbp]
    \centering
    \caption{Characteristic of our framework compared to SoA alternatives}
    \vspace{-1em}
    \begin{tabular}{lll}
        \toprule
        \textbf{Works} & \textbf{SW Scope} & \textbf{HW requirements for control flow management} \\
        \midrule
        EffiMap~\cite{pc_control1}    & CDFG    &bi-directional jumps,  Constant register files\\
        Marionette~\cite{control_plane} & CDFG    & Reconfiguration unit, CFG control network \\
        SatMapIt~\cite{satjournal}   & DFG     & External CPU for CFG management \\
        \textbf{Ours}       & CDFG    & / \\
        \bottomrule
    \end{tabular}
    \label{tab:hardware}
    \vspace{-1em}
\end{table}

\begin{table}[htbp]
    \centering
    \caption{Benchmark Categories and Attributes}
    \vspace{-1em}
    \begin{tabular}{llp{5cm}p{4cm}}
        \toprule
        \textbf{Types} & \textbf{Benchmarks} & \textbf{CFG Attributes} & \textbf{I/O Data sizes} \\
        \midrule
        \multirow{3}{*}{Convolution} 
            & conv3d  & 6-level nested loops &  3x15x15 input,  2x5x5 kernel  \\ 
            & conv2d  & 4-level nested loops &  12x12 input,  3x3 kernel  \\
            & conv1d  & 3-level nested loops &  1x64, input  1x3 kernel \\
        \midrule
        \multirow{5}{*}{Linear Algebra} 
            & bicg    & 1-level+ 2-level nested loops & 20 x 30 data \\
            & gemm    & 4-level nested loops with control & 3 10x10 matrix   \\
            & symm    & 3-level nested loops with controls &  3 10x10 matrix  \\
            & 2mm     & 2 sequential 3-level nested loops &  4 8x8 matrix \\
            & 3mm     &  3 sequential 3-level nested loops &  7 {12x12 matrix} \\
        \midrule
        \multirow{3}{*}{Signal Processing} 
            & fir & 2-level nested Loops with branches & 1x16 input,  1x5 coefficients \\
            & gsm & Loop with multiple branches & 1x50 arrays \\
            & sha\_transform & 4 sequential compute intensive loops &  1x16 data,  1x80  weight\\
        \bottomrule
    \end{tabular}
    \label{tab:benchmarks}
    \vspace{-1em}
\end{table}

\textbf{SoA methods.} We compare our method and target platform with state-of-the-art mapping methods and their respective hardware requirements in Table \ref{tab:hardware}. EffiMap~\cite{pc_control1} supports CDFG mappings with PC-control, while bi-directional jumps are implemented by a dedicated control unit.
Constant register files are utilized to generate and manage constant values.
Conversely, we achieve these functionalities entirely through software techniques, as explained in Section 5.3.
A bi-directional jump is replaced with a one-direction conditional jump, the false case is executed with an unconditional jump, and constants are generated through combinations of arithmetic operations. 
Marionette~\cite{control_plane} also supports CDFG but requires complex hardware components, including a reconfiguration unit and a CFG control network to propagate the DFG switch decision.
SatMapIt~\cite{satjournal} focuses only on DFG mapping, leveraging external CPUs for CFG management. We consider this methodology both as a stand-alone baseline strategy and as the modulo scheduler employed in our framework.

\textbf{Benchmarks.} We consider a representative set of benchmarks including convolution for deep learning, linear algebra from PolyBench~\cite{pouchet2012polybench}, and signal processing from MiBench~\cite{Mibench}. 
These benchmarks include explicit address computations and conditional execution divergence, allowing us to evaluate the compiler on applications with dynamic execution patterns.
We use modest data sizes to enable fair comparison with prior work and to isolate the impact of our CDFG-based scheduling on data movement and control-flow handling. We focus on preserving general application behavior, including dynamic and data-dependent execution patterns, rather than applying aggressive loop transformations. Exploring MLIR-based optimizations to further expose spatial parallelism is left for future work.
The benchmarks, along with their data sizes and CFG attributes, are detailed in Table \ref{tab:benchmarks}.

\section{Results}
\label{sec:res}
We evaluate our approach from multiple points of view. First, we demonstrate that our CDFG stance effectively increases the compilation scope with respect to DFG-based alternatives. Then, we report the effect of the devised CFG-based optimizations on compile time. Furthermore, we discuss the run-time performance improvements that result from each optimization technique introduced in this work. Finally, we compare our compilation framework with state-of-the-art approaches and their respective target platforms, demonstrating that our method offers higher mapping efficiency while requiring minimal hardware for controlling execution.

\subsection{IR scope analysis}

\begin{figure}[b]
    \centering
    \includegraphics[width=1\linewidth]{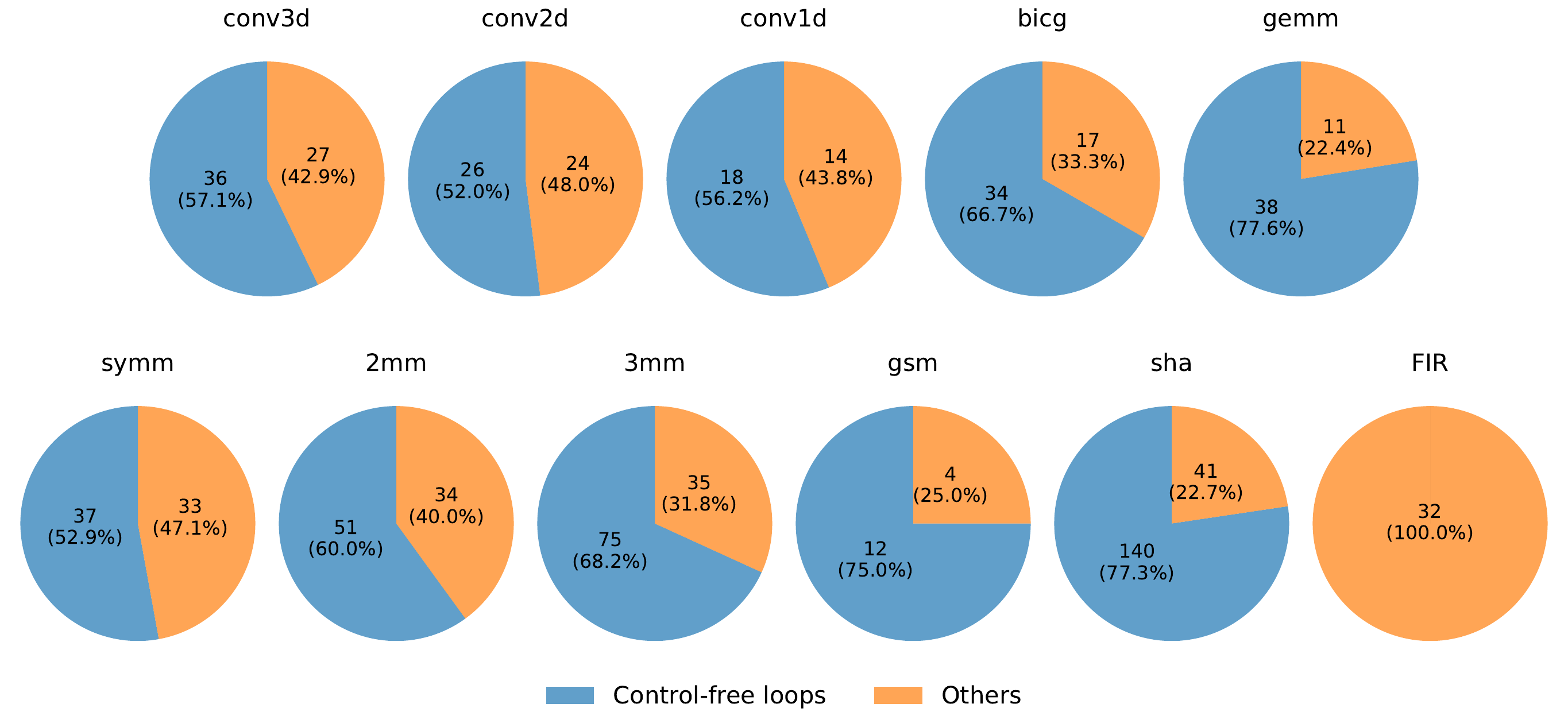}
    \caption{Break-down of operations in the considered benchmarks, reporting the percentage of operations belonging to the most internal loops. Only those can be mapped on CGRA using DFG-only strategies, while our approach allows for the entirety of kernels to be CGRA-accelerated. }
    \label{fig:res scope}
\end{figure}
Figure \ref{fig:res scope} illustrates the applicable scope of DFG mappers, where the blue sections represent operations for control-free loops and the orange sections indicate operations outside the innermost loops, for which control flow considerations are required.
The numbers and ratios shown in the pie charts indicate the count of operations and their proportion relative to the total number of operations in each application.
{The results indicate that control-free loops account for only a portion of the benchmarks, and that considering control flow can effectively increase the scope of CGRA kernels that can be compiled.}
The CFG of \texttt{FIR} cannot be simplified to include a control-free loop that is targetable by DFG mappers, which must be mapped by a CDFG compiler.
{Moreover, DFG-only approaches require the orchestration of CGRA configurations across multiple control-free loops, as well as frequent input/output data transfers between the CGRA and the host processor, resulting in significant overheads, as discussed in Section 9.4.}
Such runtime overheads are entirely waived in our proposed strategy, as a single configuration is required to support the acceleration of the entirety of each kernel.

\subsection{Compilation time}
\label{sec: res compile time}

\begin{figure}[t!]
    \centering
    \begin{subfigure}[b]{0.49\linewidth}
        \centering
        \includegraphics[width=\linewidth]{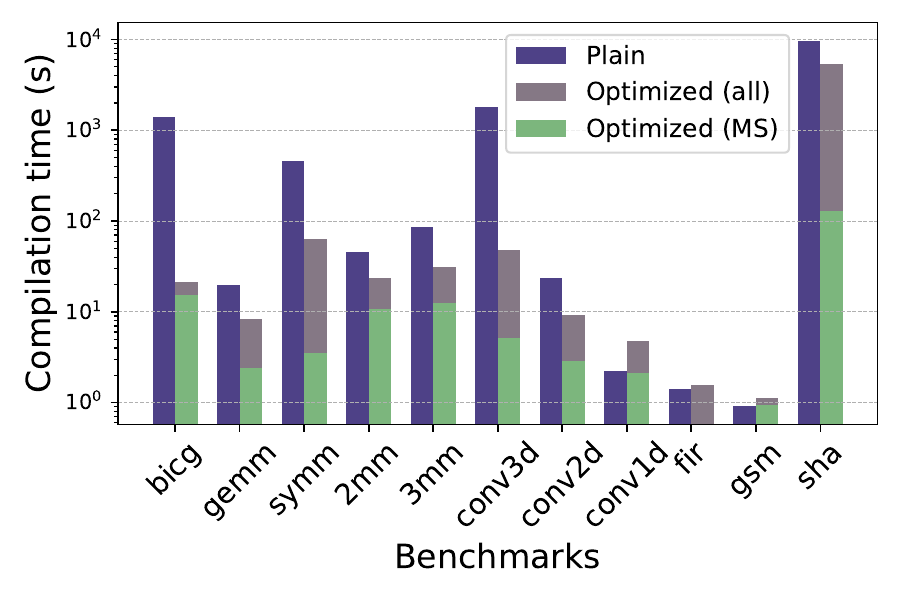}
        \caption{Compilation for 3x3 CGRA}
    \end{subfigure}
    \begin{subfigure}[b]{0.49\linewidth}
        \centering
        \includegraphics[width=\linewidth]{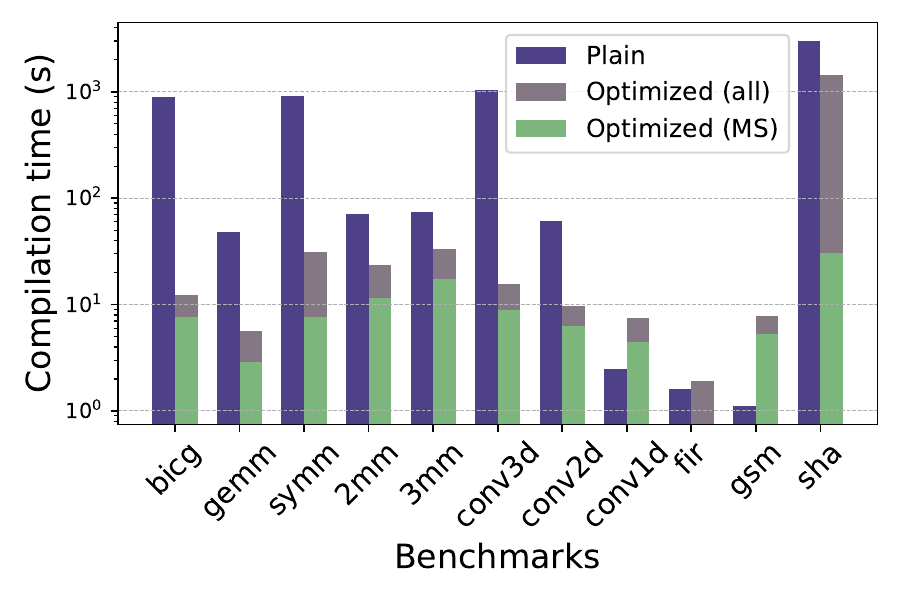}
        \caption{Compilation for 4x4 CGRA}
    \end{subfigure}
    \caption{{Compilation time comparison on Intel(R) i7-13700H. Plain indicates mapping for IR generated by MLIR front end, and Optimize shows the compilation time where the IR undergoes all optimizations supported in the framework. Lower values are better.}}
    \label{fig:compile_time}
\end{figure}
The compilation time encompasses the whole compilation process required to map an application, including the front, middle, and back end for IR transformation, operation mapping, and register allocation for assembly generation. Figure~\ref{fig:compile_time} illustrates the total compilation time targeting 3×3 and 4×4 CGRAs.
In the plain compilation flow, the operation mapping is performed directly on the IR generated by the MLIR front end. 
In contrast, the optimized flow processes the IR through CFG simplification (Section~\ref{sec:control optim}), {then applies MS to map control-free loops, followed by CDFG mapping, which takes the MS result as input and performs final placement and routing (Section~\ref{sec:optimization}).}
{Comparing 3×3 and 4×4 CGRAs, a larger array does not necessarily lead to longer compilation time. While a 4×4 CGRA enlarges the scheduling and routing search space, thereby increasing mapping complexity, it can also provide additional spatial resources that help resolve resource conflicts more efficiently, as observed in benchmarks such as \texttt{bicg}.}
CDFG mapping is decomposed into the mapping of its basic constituent blocks in our methodology. 
Therefore, the total mapping time is mainly influenced by two factors: the number of operations within individual basic blocks and the total number of basic blocks. Mapping complexity increases exponentially with the number of operations in a basic block, making blocks with large DFG sizes the primary contributors to overall mapping time. Additionally, the presence of multiple basic blocks with high mapping complexity further extends the compilation time. For example, the \texttt{gsm} benchmark contains six basic blocks, each with fewer than five operations, resulting in a relatively fast compilation. In contrast, \texttt{sha} comprises eight basic blocks, each containing more than 25 operations, significantly increasing the mapping time.

Operation mapping dominates the overall compilation time, whereas the time spent on IR transformation and register allocation is negligible. 
This is exemplified by the \texttt{FIR} benchmark, whose CFG cannot be simplified by our approach. 
Hence, the observed increase in compilation time is attributed solely to IR transformation, which contributes only a few milliseconds.
The optimized implementation exhibits longer compilation times compared to the plain version for the \texttt{conv1d} benchmark, with compilation overhead increasing by several seconds.  In the case of conv1d, the control-free kernels contain small counts of operations with reduced dependencies, making the plain version easier to find a solution.
The modulo scheduler performs more extensive exploration to identify initiation intervals that can exploit the available parallelism. 
In fact, most optimized compilation flow achieves faster compilation than the plain methodology. 
This improvement comes from CFG simplification, which restructures the CFG to be amenable to modulo scheduling. The SAT-based modulo scheduler employed in our flow reduces mapping time for computation-intensive loop kernels.

\subsection{Impact of optimizations on runtime}
Three factors influence the runtime efficiency of the mapping results produced by the proposed framework, ranked in order of their impact:
(1) the register attributes used to assign inter-basic-block values, as discussed in Section~\ref{sec:in or out};
(2) CFG simplification techniques that change the CFG structure, detailed in Section~\ref{sec:control optim}; and
(3) transformations based on modulo scheduling to enable loop-level parallelism, described in Section~\ref{sec:optimization}.
Figure \ref{fig:combined} illustrates the performance improvements achieved through different combinations of optimizations.
Plain and Optimized compilation settings are defined as  in Section~\ref{sec: res compile time}— in Plain no optimization is applied, while in Optimized all are employed. 
Additionally, we evaluate intermediate configurations that (1) apply only value allocations in both internal and external registers, and (2) apply CFG simplifications further.

\begin{figure}[t!]
    \centering
    \begin{subfigure}[b]{1\textwidth}
        \centering
        \includegraphics[width=0.9\linewidth]{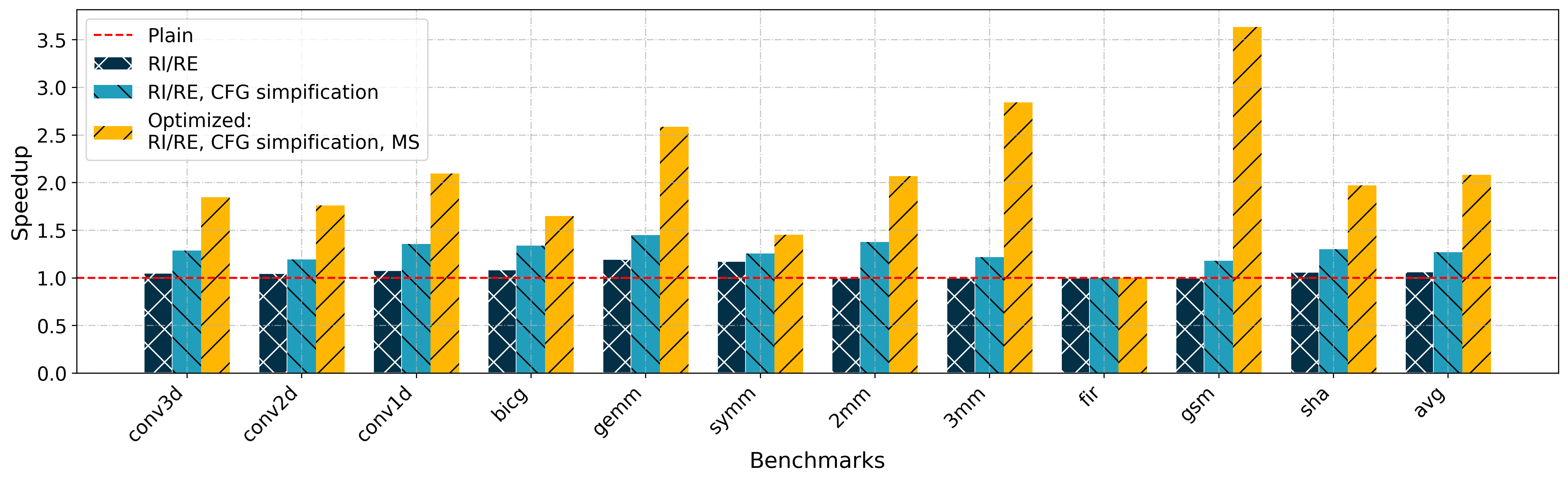}
        \caption{{Runtime comparison: Applying optimizations on a 3x3 CGRA. }}
        \label{fig:res 3x3}
    \end{subfigure}%
    \hfill
    \begin{subfigure}[b]{1\textwidth}
        \centering
        \includegraphics[width=.9\linewidth]{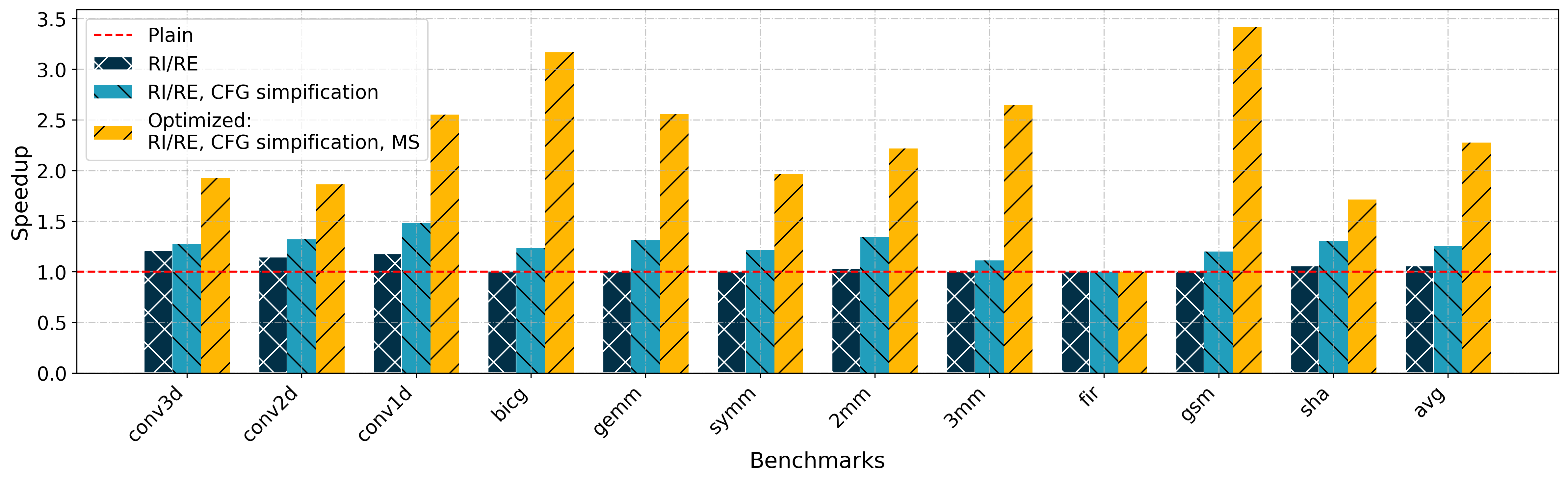}
        \caption{{Runtime comparison: Applying optimizations on a 4x4 CGRA.}}
        \label{fig:res 4x4}
    \end{subfigure}
    \caption{Impact of register allocation for live values, CFG simplification, and modulo scheduling across various CGRA sizes. RI/RE denotes our optimized Register Internal/External placement strategy. The use of CFG simplification and modulo scheduling is indicated when the respective technique is applied. Higher values rare better.}
    \label{fig:combined}
    \vspace{-1.5em}
\end{figure}

The placement of values propagated across basic blocks significantly impacts convolution and linear algebra benchmarks, as these workloads carry high data dependencies through nested loops for computing memory addresses. For a 4×4 CGRA running convolutional benchmarks, selectively placing partial variables in external registers expands the access range, resulting in an 6.6\% acceleration gain on average.
Combinational use of CFG transformations simplifies control flow, particularly in \texttt{gsm}, where five blocks are merged into a single self-iterating loop.
This CFG simplification enables multiple blocks to be consolidated, allowing merged operations to be scheduled simultaneously.
As a result, acceleration gains of 26.8\% and 26.4\% are achieved on 3×3 and 4×4 CGRAs, respectively. For \texttt{conv1d} on a 4×4 CGRA and \texttt{gemm} on a 3×3 CGRA, performance improvements reach up to 48\% and 45\%, respectively.
Loops with lower nested levels such as \texttt{gsm} and \texttt{gemm} obtain 316\% and 255\% runtime faster on a 4x4 CGRA.
On average, modulo scheduling folds the innermost loop, achieving 197\% on 3x3 and 223\% acceleration on 4x4 CGRAs, indicating $L \approx 2*II$ achieved by the modulo scheduler.
CGRAs with larger grid sizes achieve higher speedups, as the presence of more hardware resources enables greater parallelism.
In general, the introduced optimization is highly beneficial on all considered benchmarks. 
\texttt{FIR} is an outlier, as its CDFG structure presents a write operations in one   branch inside an innermost loop, which disallow  basic block merging.

\subsection{Comparison with state-of-the-art methodologies}
\label{sec: compare soa}

\begin{figure}[t]
    \centering
    \includegraphics[width=1\linewidth]{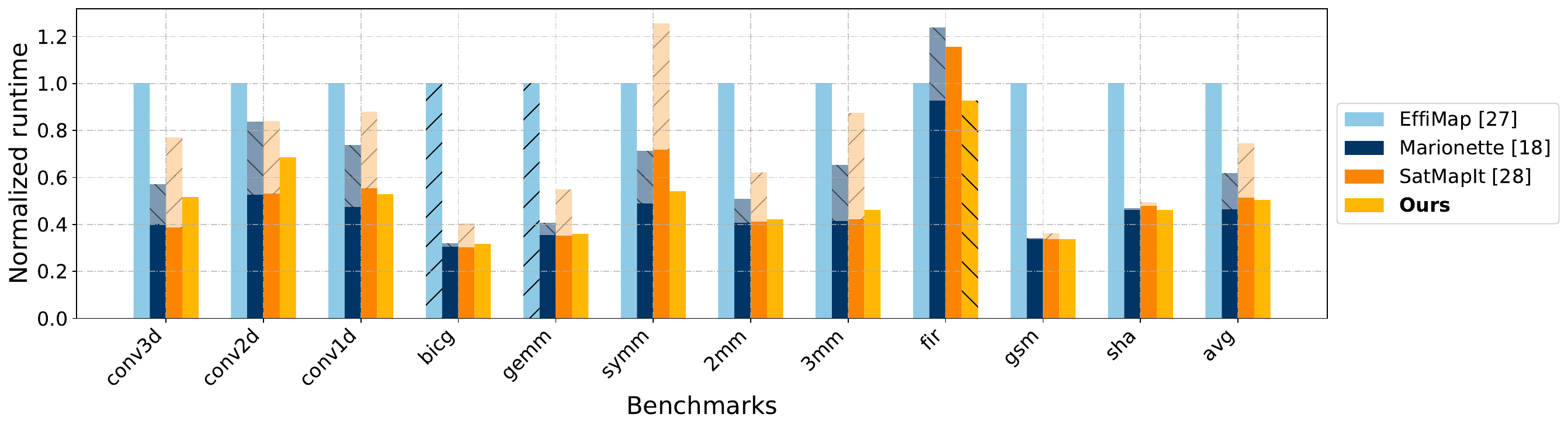}
    \caption{{Evaluation of normalized runtime across different architectures. EffiMap and our method employ a single CGRA configuration for supporting entire kernels, while Marionette and SatMapIt require reconfiguration during basic block switches, with the reconfiguration overhead shown as a semi-transparent color. Lower values are better.}}
    \label{fig:res_soa}
    \vspace{-1em}
\end{figure}

Figure \ref{fig:res_soa} shows the execution runtime of benchmarks on  4x4 CGRAs  for the methodologies described in Table \ref{tab:hardware}.
Results are normalized with respect to EffiMap~\cite{pc_control_mapping}, which supports CDFG mapping without reconfiguration overhead.
Marionette~\cite{control_plane} also supports CDFG mapping but involves reconfiguration time for basic block switching, depicted in semi-transparent dark blue.
Since in the case of SatMapIt~\cite{satjournal} only a portion of benchmarks can be CGRA-accelerated, we optimistically assume that non-CGRA code runs on a processor with an IPC (instructions-per-cycle) equal to 1. In Figure \ref{fig:res_soa}, non-CGRA run time plus CGRA configuration overhead for SatMapIt is   depicted in semi-transparent orange.

Our compilation approach achieves the lowest compilation runtime across all benchmarks, as shown in Figure \ref{fig:res_soa}. We achieve average speedups of 2.12×, 1.48×, and 1.20× over EffiMap~\cite{pc_control_mapping}, SAT-MapIt~\cite{satjournal}, and Marionette~\cite{control_plane}, respectively.
EffiMap does not fully exploit the optimization spaces during the compilation process: it relies on dedicated hardware to support instruction-level parallelism, but neglects  loop-level parallelism opportunities.
However, loop-level parallelism is critical for optimizing runtime efficiency for CGRA, as discussed in Section 9.3, where modulo scheduling achieves $L\approx 2*II$.
Our CFG adaptation creates new kernels with loop length $L'=II\approx 0.5*L$ on average.
EffiMap achieves faster runtime for \texttt{FIR} because modulo scheduling cannot be performed due to the presence of non-mergeable branches inside an innermost loop.
Our compilation method outperforms theirs because their heuristic strategy does not optimize for the shortest basic block runtime, leading to suboptimal solutions.

When the computational workloads dominate the benchmarks with control-free, non-nested loops, both SATMapIt and Marionette achieve performance comparable to our method, as the runtime reconfiguration overhead becomes negligible. This is shown in benchmarks such as \texttt{gsm} and \texttt{sha}.
However, both SatMapIt and Marionette experience significant reconfiguration overhead for nested loops. For example, a \texttt{conv3d} application with an input tensor of dimensions $D \times H \times W$ (depth × height × width) and a kernel of dimensions $d \times h \times w$ would require reconfiguration $D \times H \times W \times d \times h$ times, the number of branches to execute the innermost loop over the width of the kernel $w$.
Even if the reconfiguration time is optimized to take only a few clock cycles, the large number of reconfigurations still introduces a substantial overhead.
For \texttt{conv2d} and \texttt{conv3d}, the computation time is faster for Marionette and SatMapIt than for our approach, as in these methods, no live-in / live-out values must be maintained across basic blocks. Nevertheless, our approach still outperforms both when reconfiguration time is accounted for, because the entire benchmark can be mapped as a single computational kernel. 

In conclusion, our compilation framework outperforms methodologies that rely on hardware reconfiguration units by eliminating the need for reconfiguration during basic block transitions. It also surpasses other PC-based control models through CFG reshaping, enabling effective support for modulo scheduling.


\section{Conclusion}

By combining efficiency and reprogrammability, CGRAs can support acceleration of computationally intensive edge applications. The key to their mainstream adoption is the availability of a compilation framework that enables the automatization of the deployment of complex compute kernels on available hardware resources.

Against this backdrop, we have proposed a novel end-to-end framework in this paper that, by leveraging and extending the MLIR infrastructure, can generate CGRA assembly from source code. We broadened the scope of CGRA scheduling by considering both control and data flows during compilation, demonstrating how this approach can efficiently compile kernels. 
Moreover, our framework supports the compilation for applications with arbitrary control flows without dedicated hardware support.
The introduced CFG-based optimizations lead to speedups of 1.96x and 2.23x on 3x3 and 4x4 CGRAs, respectively, resulting in, on average, 2.12x faster execution than state-of-the-art methods.


\section*{Acknowledgment}
This work has been partially supported by the
Swiss NSF Edge-Companions project (GA No. 10002812) and by the Swiss State Secretariat for Education, Research, and Innovation (SERI) through the SwissChips research project. This research was partially conducted by ACCESS – AI
Chip Center for Emerging Smart Systems, supported by the InnoHK initiative of the Innovation and Technology Commission of the Hong Kong Special Administrative Region Government.
This work was supported by the Swiss National Science Foundation via project ADApprox (grant 200020\_188613).

\bibliographystyle{IEEEtran}
\bibliography{reference}

\end{document}